\documentclass[11pt]{revtex4}
\usepackage{mathtools}
\usepackage{hyperref}
\usepackage{pdflscape}
\usepackage{graphics}
\usepackage{subfigure}
\usepackage{epsfig}
\usepackage{amsmath,amsfonts}
\usepackage{amssymb}
\usepackage{color}
\usepackage{ulem}
\usepackage{graphicx}
\begin{document}
\title{Polarization modes of gravitational waves in general modified gravity: general metric theory and general scalar-tensor theory}

\author{Yu-Qi Dong$^{a,b,c}$}
\email{dongyq2023@lzu.edu.cn}

\author{Yu-Qiang Liu$^{a,b,c}$}
\email{liuyq18@lzu.edu.cn}

\author{Yu-Xiao Liu$^{a,b,c}$}
\email{liuyx@lzu.edu.cn (corresponding author)}

\affiliation{$^{a}$ Institute of Theoretical Physics \& Research Center of Gravitation, Lanzhou University, Lanzhou 730000, China\\
	$^{b}$ Key Laboratory of Quantum Theory and Applications of MoE, Lanzhou University, Lanzhou 730000, China\\
	$^{c}$ Lanzhou Center for Theoretical Physics \& Key Laboratory of Theoretical Physics of Gansu Province, Lanzhou University, Lanzhou 730000, China}

\begin{abstract}
\textbf{Abstract:} In this paper, we establish a unified parameterized framework for analyzing the polarization modes of gravitational waves in the general metric theory (where gravity is only described by the metric) and the general scalar-tensor theory (where gravity is described by the metric and an additional scalar field). Specifically, we study the polarization modes of gravitational waves in the most general metric theory and general scalar-tensor theory that satisfy the following conditions: (1) Spacetime is four-dimensional; (2) The theory satisfies the principle of least action; (3) The theory is generally covariant; (4) The action describing a free particle is $\int ds$. We find that the polarization modes of gravitational waves in the theory satisfying the above conditions depend on the selection of parameters in the framework, and the theory allows for up to all six polarization modes. Once we have established our framework, the analysis of the polarizations of gravitational waves in specific theories will depend on determining the corresponding parameters within our framework.

We also find that the polarization modes of gravitational waves in the general metric theory and the general scalar-tensor theory that satisfy the conditions also have some interesting universal properties. 

\end{abstract}
	
\maketitle

\section{Introduction}
\label{sec: intro}
Polarization is a fundamental property of gravitational waves. General relativity only allows two independent polarization modes of gravitational waves \cite{MTW}. They are two tensor modes, namely the $+$ mode and the $\times$ mode. However, gravitational waves in the four-dimensional modified gravity theory have up to six independent polarization modes \cite{Eardley}. For a specific modified gravity, the field equation will constrain the number of polarization modes. Therefore, it usually only allows a subset of six modes. Different modified gravity theories have different predictions for polarization modes of gravitational waves. Moreover, these theories also have different predictions for the speed of the corresponding polarized gravitational waves. Therefore, the detection of polarization modes of gravitational waves is crucial for testing various modified gravity theories.

Gravitational waves have been successfully detected \cite{Abbott1,Abbott2,Abbott3,Abbott4,Abbott5}. Various gravitational wave detection projects are also continuously planned and executed worldwide. This makes it possible to detect the polarization modes of gravitational waves. For instance, ground-based gravitational wave detectors, such as LIGO, VIRGO, and KAGRA, can detect gravitational waves in different two-dimensional spatial planes. Consequently, they can collaborate to detect gravitational wave polarization modes \cite{Abbott4,Abbott5,H. Takeda,B. P . Abbottet al.1,B. P . Abbottet al.2,B. P . Abbottet al.3,Atsushi Nishizawa,Kazuhiro Hayama,Maximiliano Isi1,Maximiliano Isi2,K. Chatziioannou,Hiroki Takeda,Yuki Hagihara,Peter T. H. Pang,B. P . Abbottet al.4}. Furthermore, in addition to ground-based gravitational wave detectors, there are expectations to use space-borne gravitational wave detectors like Lisa, Taiji, and TianQin {\cite{lisa,taiji,Lisa-taiji,tianqin,tianqin2,tianqin3}} in the future to detect gravitational wave polarization modes. These space-borne gravitational wave detectors are already in organized preparation and are scheduled to commence operations around 2035 \cite{Y.M.Hu}. Finally, there is a potential to utilize Pulsar Timing Arrays (PTA) \cite{YiPeng Jing} for the detection of gravitational wave polarization modes \cite{Z. Chen,Z. Chen2,G. Agazie}.

Currently, numerous studies have explored the polarization modes of gravitational waves within specific modified gravity theories, such as $f(R)$, Horndeski, generalized Proca,  dynamical Chern-Simons, Palatini-Horndeski, Einstein-dilaton-Gauss-Bonnet, $f(T)$, teleparallel Horndeski, Bumblebee, Horava, Palatini generalized Brans-Dicke, Einstein-aether, tensor-vector-scalar, Chern-Simons Axion $f(R)$, and scalar-tensor-vector theories \cite{f(R,f(R2,Horndeski0,TeVeS,Horava,STVG,fT,dCS and EdGB,Y.Dong,L.Shao,S. Bahamonde,J.Lu,Y.Dong2,S. Nojiri}. There are also studies on methods for analyzing polarization modes of gravitational waves \cite{Eardley,N-P,TH. Hyun,Kristen Schumacher,M.Alves}. In addition, Ref. \cite{Y.Liu} studied the polarization modes of gravitational waves with extra spatial dimensions. References \cite{S. Nojiri2,S. Nojiri3} studied the production and propagation of gravitational waves in modified gravity in the context of cosmology. References \cite{E. Battista,E. Battista2} studied gravitational radiation in modified gravity. Upon reviewing existing studies on the polarization modes of gravitational waves in specific gravity theories, one can observe that while different theories yield various quantitative predictions for the polarization modes of gravitational waves, they share some commonalities in certain aspects. For example, there are tensor modes propagating at the speed of light and there are no vector modes in $f(R)$ theory, Horndeski theory and Palatini-Horndeski theory \cite{f(R,Horndeski0,Y.Dong}.
In addition, when the mass of a scalar mode in these theories is not zero, this mode is a mixed mode of the longitudinal mode and breathing mode. And when the mass of the scalar mode is zero, it becomes a breathing mode. Another example is Einstein-aether theory and generalized Proca theory \cite{TeVeS,Y.Dong2}. They all do not allow vector gravitational waves only when tensor gravitational waves propagate at the speed of light. These shared characteristics among polarization modes of gravitational waves in different theories naturally lead to the following questions. What are the premises and reasons for the establishment of these commonalities? And how can we accurately express these commonalities in the form of theorems?

We hope to answer the above questions by establishing a unified parameterized framework for analyzing the polarization modes of gravitational waves. This framework can encompass various modified gravity theories that satisfy certain predetermined conditions. The most direct way to establish this framework seems to be to write the most general action that satisfies the given conditions, and then to study the polarization modes of gravitational waves with this action. We can obtain the general properties of the polarization mode of gravitational waves under these conditions. It can be imagined that the polarization modes of gravitational waves obtained in this manner will depend on the selection of certain parameters. These parameters are essentially derived from the most general action and the determination of the polarization modes of gravitational waves in a specific theory satisfying these conditions is now completely equivalent to determining these parameters.

However, writing the most general action is often very difficult. This road seems impassable. But it does not mean that we can not establish a unified framework. In fact, we notice that since gravitational waves are very weak, we only need to analyze the linearized field equations. That is to say, in order to study the polarization modes of gravitational waves, it is only necessary to know the second-order term of the perturbation expansion of the action. Therefore, in order to establish a unified parameterization framework, we do not need to write the complete action, but just need the most general second-order action for perturbations. Writing the most general second-order action that satisfies certain conditions is relatively easy and generally possible. This is also the method by which we establish a unified framework in this paper.

The general metric theory and the general scalar-tensor theory are two main types of modified gravity theories. In the general metric theory, gravity is only described by the metric $g_{\mu\nu}$. While in the general scalar-tensor theory, gravity is described by the metric $g_{\mu\nu}$ and an additional scalar field $\Psi$. The goal of this paper is to establish a unified parameterized framework for analyzing the polarization modes of gravitational waves in the general metric theory and the general scalar-tensor theory that satisfy the following conditions: (1) Spacetime is four-dimensional; (2) The theory satisfies the principle of least action; (3) The theory is generally covariant; and (4) The action describing a free particle is given by $\int ds$. Therefore, in Sec. \ref{sec: 2}, we construct the most general second-order action of general metric theory that satisfies the above conditions. 
In Sec. \ref{sec: 3}, we analyze the polarization modes of gravitational waves for the most general second-order action of the general metric theory. In Sec. \ref{sec: 4}, we obtain some universal properties of the polarization modes of gravitational waves in the general metric theory. Similarly, in Sec. \ref{sec: 5}, we construct the most general second-order action of the  general scalar-tensor theory that satisfies the above conditions. In Sec. \ref{sec: 6}, we analyze the polarization modes of gravitational waves for the most general second-order action of the general scalar-tensor theory and 
obtain some universal properties of the polarization modes. Finally, Sec. \ref{sec: 7} is the conclusion.

We use $c=G=1$ and the metric signature $(-,+,+,+)$. The indices $(\mu,\nu,\lambda,\rho)$ range over four-dimensional spacetime indices ($0,1,2,3$), and the indices $(i,j,k,l)$ range over three-dimensional spatial indices ($1,2,3$), which correspond to ($+x,+y,+z$) directions, respectively.

\section{Second-order action of the general metric theory}
\label{sec: 2}
As mentioned in the previous section, to establish a unified framework for studying the polarization modes of gravitational waves in the most general four-dimensional metric theory, there is no need to write down the most general action. We just need to construct the most general action of second-order perturbations, which will be done in this section.

We assume that the Minkowski metric $\eta_{\mu\nu}$ is a solution of the general metric theory and consider a perturbation $h_{\mu\nu}$ on the background metric $\eta_{\mu\nu}$:
\begin{eqnarray}
	\label{perturbation}
	g_{\mu\nu}=\eta_{\mu\nu}+h_{\mu\nu},\quad \left|h_{\mu\nu}\right|\ll1.
\end{eqnarray}
In this manner, the second-order action of the general metric theory will only be a functional of $h_{\mu\nu}$. In the following, $\eta_{\mu\nu}$ and $\eta^{\mu\nu}$ are used to lower and raise the four-dimensional spacetime index and $h$ is defined as $\eta_{\mu\nu}h^{\mu\nu}$.

The conditions from the previous section that we apply to the general metric theory will impose the following requirements on the second-order action:
\begin{itemize}
	\item [(1)] Each term in the action is a second order of $h_{\mu\nu}$.
	\item [(2)] The second-order action is invariant under the gauge transformation $x^{\mu} \rightarrow x^{\mu}+\xi^{\mu}(x)$.
\end{itemize}
In the previous section, the condition that the theory satisfies the principle of least action tells us to only study the general metric theory whose action can be given. Since we only have the metric perturbation $h_{\mu\nu}$ as a variable, each term of the second-order action must only have two $h_{\mu\nu}$, and we call it in the form of $hh$. This is the requirement of the condition (2) here. Finally, the condition (3) is a requirement for generally covariant. If a theory is generally covariant, then the linearized version of the theory will satisfy gauge symmetry \cite{Michele Maggiore}.

In this paper, we consider that spacetime is described by Riemannian geometry. For Riemannian geometry, the only intrinsic quantities in its algebraic structure are the background metric $\eta_{\mu\nu}$ and the perturbation $h_{\mu\nu}$ given by the inner product structure of the tangent space, the four-dimensional Levi-Civita totally antisymmetric tensor $E_{\mu\nu\lambda\rho}$ given by the exterior product structure of the tangent space, and the partial derivative $\partial_{\mu}$ given by the differential structure. (Reference \cite{Lavinia Heisenberg} provides a physical introduction to various structures on manifolds.) The action should be represented by the intrinsic quantities mentioned above. If another tensor appears in the action, we need to explain the physics of the new tensor. Guided by the concept that classical gravity is described by geometry, this new tensor requires us to add a new algebraic structure to the current manifold. For the above considerations, we also require
\begin{itemize}
	\item [(4)] Each term in the second-order action can be represented as a combination of $\eta_{\mu\nu}$, $\partial_{\mu}$, $E_{\mu\nu\lambda\rho}$, $h_{\mu\nu}$ and the coupling parameters of the theory.
\end{itemize}
It should be pointed out that the most general second-order action does not actually include $E_{\mu\nu\lambda\rho}$. This can be seen from the following analysis. Firstly, we can always consider only the case where the term in the action contains only one $E_{\mu\nu\lambda\rho}$ without losing generality. This is because we have \cite{landau}
\begin{eqnarray}
	\label{EE=delta}
	E^{\alpha\beta\gamma\sigma}E_{\mu\nu\lambda\rho}
	~=~-
	\begin{vmatrix}
			~\delta^{\alpha}_{\mu}~ & \delta^{\alpha}_{\nu}~ & \delta^{\alpha}_{\lambda}~ & \delta^{\alpha}_{\rho}~
			\\
			~\delta^{\beta}_{\mu}~ & \delta^{\beta}_{\nu}~ & \delta^{\beta}_{\lambda}~ & \delta^{\beta}_{\rho}~
			\\
			~\delta^{\gamma}_{\mu}~ & \delta^{\gamma}_{\nu}~ & \delta^{\gamma}_{\lambda}~ & \delta^{\gamma}_{\rho}~
			\\
			~\delta^{\sigma}_{\mu}~ & \delta^{\sigma}_{\nu}~ & \delta^{\sigma}_{\lambda}~ & \delta^{\sigma}_{\rho}~
	\end{vmatrix}.
\end{eqnarray}
Here, $\delta^{\mu}_{\nu}$ is the Kronecker delta. Therefore, trems with odd numbers of $E_{\mu\nu\lambda\rho}$ can always be written in the form of only one $E_{\mu\nu\lambda\rho}$, while terms with even numbers of $E_{\mu\nu\lambda\rho}$ can always be written without any $E_{\mu\nu\lambda\rho}$. The problem now boils down to the case where there is only one $E_{\mu\nu\lambda\rho}$. Just note that $\eta_{\mu\nu}$ and $h_{\mu\nu}$ are symmetric with respect to indices $(\mu, \nu)$, and it can be found that terms containing only one $E_{\mu\nu\lambda\rho}$ in the action must be equal to $0$. This completes our analysis. Therefore, the term in the action will only be a combination of $\eta_{\mu\nu}$, $\partial_{\mu}$, $h_{\mu\nu}$ and the coupling parameters.

Using the condition (4), we can obtain an important result: terms of the field equation in the general metric theory only has an even number of derivatives. This is because $\eta_{\mu\nu}$ and $h_{\mu\nu}$ are second-order tensors, so a tensor formed by combining $\eta_{\mu\nu}$ and $h_{\mu\nu}$ must have an even number of indices. And we cannot form a scalar with an odd number of $\partial_{\mu}$ and the two tensors $\eta_{\mu\nu}$ and $h_{\mu\nu}$.

Therefore, we only need to consider the most general second-order action that can derive the field equation with up to $2N$-th derivative terms. In order for the second-order action to satisfy the conditions (1), (2), and (4) in this section, it should have the following form:
\begin{eqnarray}
	\label{S=Sigma SI metric}
	S=\sum_{I=0}^N S_{I},
\end{eqnarray}
where $S_{I}$ is the most general action that can derive all $2I$-th derivative terms in the field equation. When $I\geq2$, it has the specific form
\begin{eqnarray}
	\label{SI even metric}
	S_{I}&=&\int d^4{x}
	\left[
	a^{(I)}_{1} \left(
	\Box^{I-2}\partial_{\mu}\partial_{\nu}\partial^{\lambda}\partial^{\rho}h_{\lambda\rho}
	\right)h^{\mu\nu}
	+ a^{(I)}_{2} \left(
	\Box^{I-1}\partial_{\nu}\partial^{\lambda}h_{\mu\lambda}
	\right)h^{\mu\nu}
	\right.
	\nonumber \\
	&+&\left.a^{(I)}_{3} \left(
	\Box^{I}h_{\mu\nu}
	\right)h^{\mu\nu}
	+a^{(I)}_{4} \left(
	\Box^{I-1}\partial_{\mu}\partial_{\nu}h
	\right)h^{\mu\nu}
	+a^{(I)}_{5} \left(
	\Box^{I}h
	\right)h
	\right].
\end{eqnarray}
When $I=1$, it has the form
\begin{eqnarray}
	\label{S1 even metric}
	S_{1}=\int d^4{x}
	\left[
	a^{(1)}_{2} \left(
	\partial_{\nu}\partial^{\lambda}h_{\mu\lambda}
	\right)h^{\mu\nu}
	+a^{(1)}_{3} \left(
	\Box h_{\mu\nu}
	\right)h^{\mu\nu}
	+a^{(1)}_{4} \left(
	\partial_{\mu}\partial_{\nu}h
	\right)h^{\mu\nu}
	+a^{(1)}_{5} \left(
	\Box h
	\right)h
	\right].
\end{eqnarray}
And when $I=0$, it has the form
\begin{eqnarray}
	\label{S0 even metric}
	S_{0}=\int d^4{x}
	\left[
	a^{(0)}_{3}
	h_{\mu\nu}
	h^{\mu\nu}
	+a^{(0)}_{5} h^2
	\right].
\end{eqnarray}
Here, $\Box=\partial^{\mu}\partial_{\mu}$ and $\Box^{I}$ is the $I$-th power of $\Box$, $a^{(I)}_{i} (i=1, 2, 3, 4, 5)$ are constants. For convenience, we also define
\begin{eqnarray}
	\label{a1,a2,a3,a4,a5}
    a^{(0)}_{1}=a^{(0)}_{2}=a^{(0)}_{4}=a^{(1)}_{1}=0.
\end{eqnarray}

In order for our action to meet the condition (3) of this section, we also need to further constrain the  parameters $a^{(I)}_{i}$ (where $i=1, 2, 3, 4, 5$) to ensure that the action satisfies the gauge invariance. Therefore, we consider the gauge transformation
\begin{eqnarray}
	\label{gauge x->x+xi}
	x^{\mu} \rightarrow x^{\mu}+\xi^{\mu}(x).
\end{eqnarray}
Here, $\xi^{\mu}$ is an arbitrary function of space-time coordinates. $\partial_{\mu} \xi_{\nu}$ and $h_{\mu\nu}$ are of the same order of magnitude. Under transformation (\ref{gauge x->x+xi}), $h_{\mu\nu}$ transforms as follows
\begin{eqnarray}
	\label{gauge transformation h}
	h_{\mu\nu} \rightarrow	h_{\mu\nu}-\partial_{\mu}\xi_{\nu}-\partial_{\nu}\xi_{\mu}.
\end{eqnarray}
Therefore, the terms in the $S_{I}$ action (\ref{SI even metric}) should satisfy the following transformation:
\begin{eqnarray}
	\label{a1 change}
a^{(I)}_{1} \left(
               	\Box^{I-2}\partial_{\mu}\partial_{\nu}\partial^{\lambda}\partial^{\rho}h_{\lambda\rho}
	        \right)h^{\mu\nu}
	&\rightarrow &
a^{(I)}_{1} \left(
               \Box^{I-2}\partial_{\mu}\partial_{\nu}\partial^{\lambda}\partial^{\rho}h_{\lambda\rho}
            \right)h^{\mu\nu}	
   \nonumber \\
  &+&4a^{(I)}_{1} \left(
                  \Box^{I-1}\partial_{\nu}\partial^{\lambda}\partial^{\rho}h_{\lambda\rho}
                \right)\xi^{\nu}
  -4a^{(I)}_{1} \left(
                  \Box^{I}\partial_{\nu}\partial^{\rho}\xi_{\rho}
                \right)\xi^{\nu},
\end{eqnarray}
\begin{eqnarray}
	\label{a2 change}
	a^{(I)}_{2} \left(
	                 \Box^{I-1}\partial_{\nu}\partial^{\lambda}h_{\mu\lambda}
	              \right)h^{\mu\nu}
	&\rightarrow &
	a^{(I)}_{2} \left(
	                 \Box^{I-1}\partial_{\nu}\partial^{\lambda}h_{\mu\lambda}
	              \right)h^{\mu\nu}
	\nonumber \\
	&+&2 a^{(I)}_{2}\left(
                   	\Box^{I-1}\partial_{\nu}\partial^{\mu}\partial^{\lambda}h_{\mu\lambda}
	              \right)\xi^{\nu}
	+2 a^{(I)}_{2}\left(
	                 \Box^{I}\partial^{\lambda}h_{\mu\lambda}
	              \right)\xi^{\mu}
	\nonumber \\
	&-&3a^{(I)}_{2}\left(
	                 \Box^{I}\partial_{\nu}\partial^{\lambda}\xi_{\lambda}
	               \right)\xi^{\nu}
	 -a^{(I)}_{2}\left(
	                  \Box^{I+1}\xi_{\mu}
	               \right)\xi^{\mu},
\end{eqnarray}
\begin{eqnarray}
	\label{a3 change}
	a^{(I)}_{3} \left(
	              \Box^{I}h_{\mu\nu}
	            \right)h^{\mu\nu}
	&\rightarrow &
	a^{(I)}_{3} \left(
	              \Box^{I}h_{\mu\nu}
	            \right)h^{\mu\nu}
	\nonumber \\
	&+&4 a^{(I)}_{3} \left(
                       \Box^{I}\partial^{\mu}h_{\mu\nu}
                     \right)\xi^{\nu}
    -2 a^{(I)}_{3} \left(
                      \Box^{I+1}\xi_{\nu}
                   \right)\xi^{\nu}
   \nonumber \\
   &-&2 a^{(I)}_{3} \left(
                     \Box^{I}\partial_{\mu}\partial^{\nu}\xi_{\nu}
                   \right)\xi^{\mu},
\end{eqnarray}
\begin{eqnarray}
	\label{a4 change}
	a^{(I)}_{4} \left(
	               \Box^{I-1}\partial_{\mu}\partial_{\nu}h
	            \right)h^{\mu\nu}
	&\rightarrow &
    a^{(I)}_{4} \left(
                   \Box^{I-1}\partial_{\mu}\partial_{\nu}h
                \right)h^{\mu\nu}
	\nonumber \\
	&+&2a^{(I)}_{4} \left(
	                   \Box^{I}\partial_{\nu}h
	                \right)\xi^{\nu}
	 -2a^{(I)}_{4} \left(
	                   \Box^{I-1}\partial_{\mu}\partial_{\nu}\partial^{\lambda}\xi_{\lambda}
	               \right) h^{\mu\nu}
	\nonumber \\
	&-&4 a^{(I)}_{4} \left(
	                   \Box^{I}\partial_{\nu}\partial^{\lambda}\xi_{\lambda}
	                 \right)\xi^{\nu},
\end{eqnarray}
\begin{eqnarray}
	\label{a5 change}
	a^{(I)}_{5} \left(
	               \Box^{I}h
	            \right)h
	&\rightarrow &
	a^{(I)}_{5} \left(
	               \Box^{I}h
	            \right)h
	\nonumber \\
	&+&4a^{(I)}_{5} \left(
	                  \Box^{I}\partial_{\rho}h
	                \right)\xi^{\rho}
	-4 a^{(I)}_{5} \left(
	                  \Box^{I}\partial^{\lambda}\partial_{\rho}\xi_{\lambda}
	               \right)\xi^{\rho}.
\end{eqnarray}

Note that there can be a partial integration difference since a partial integration does not change the value of the action. The gauge invariance requires that the action remains unchanged under transformations (\ref{a1 change})-(\ref{a5 change}) \cite{Michele Maggiore}. Therefore, the following quantity should be zero:
\begin{eqnarray}
	\label{gauge delta action is 0}	
	&&
	2\left(2a^{(I)}_{1}+a^{(I)}_{2}+a^{(I)}_{4}\right)
	\left(
	   \Box^{I-1}\partial_{\nu}\partial^{\lambda}\partial^{\rho}h_{\lambda\rho}
	\right)\xi^{\nu}
	\nonumber \\
	&+&\left(-4a^{(I)}_{1}-3a^{(I)}_{2}-2a^{(I)}_{3}-4a^{(I)}_{4}-4a^{(I)}_{5}\right)                   \left(
	   \Box^{I}\partial_{\nu}\partial^{\rho}\xi_{\rho}
	  \right)\xi^{\nu}
	\nonumber \\
	&+&2\left(a^{(I)}_{2}+2a^{(I)}_{3}\right)
	 \left(
	    \Box^{I}\partial^{\lambda}h_{\mu\lambda}
	 \right)\xi^{\mu}
	 \nonumber \\
	&-&2\left(a^{(I)}_{2}+a^{(I)}_{3}\right)
	  \left(
	     \Box^{I+1}\xi_{\mu}
	  \right)\xi^{\mu}
	 \nonumber \\
	&+&2\left(a^{(I)}_{4}+2a^{(I)}_{5}\right)
	\left(
	   \Box^{I}\partial_{\rho}h
	\right)\xi^{\rho}
	=0.
\end{eqnarray}
Equation (\ref{gauge delta action is 0}) requires that the parameters in the action (\ref{SI even metric}) should satisfy
\begin{equation}
	\begin{cases}
		2a^{(I)}_{1}+a^{(I)}_{2}+a^{(I)}_{4} = 0 \\
		4a^{(I)}_{1}+3a^{(I)}_{2}+2a^{(I)}_{3}+4a^{(I)}_{4}+4a^{(I)}_{5} = 0 \\
		a^{(I)}_{2}+2a^{(I)}_{3} = 0 \\
		a^{(I)}_{4}+2a^{(I)}_{5} = 0.
	\end{cases}
\end{equation}
Therefore, for $I=0$, it can be inferred from Eq. (\ref{a1,a2,a3,a4,a5}) that
\begin{eqnarray}
	\label{a(0)is0}
	a^{(0)}_{1}=a^{(0)}_{2}=a^{(0)}_{3}=a^{(0)}_{4}=a^{(0)}_{5}=0.
\end{eqnarray}
For $I=1$, we have
\begin{eqnarray}
	\label{a(1)}
	a^{(1)}_{1}=0,~ a^{(1)}_{2}=2a^{(1)}_{5},~ a^{(1)}_{3}=-a^{(1)}_{5},~ a^{(1)}_{4}=-2a^{(1)}_{5},
\end{eqnarray}
which shows that there is only one free parameter. For $I\geq2$, we have two free parameters:
\begin{eqnarray}
	\label{a(I)geq2}
	a^{(I)}_{1}=a^{(I)}_{3}+a^{(I)}_{5},~ a^{(I)}_{2}=-2a^{(I)}_{3},~ a^{(I)}_{4}=-2a^{(I)}_{5}.
\end{eqnarray}

We redefine $a^{(I)}_{5}$ as $a_{I}$ and $a^{(I)}_{3}$ as $b_{I}$. In this way, considering the gauge symmetry, the most general second-order action that can derive the field equation with up to $2N$-th derivative terms can be written as
\begin{eqnarray}
	\label{S gauge sym metric}
	S=S_1+\sum_{I\geq2}^N S_{I},
\end{eqnarray}
where
\begin{eqnarray}
	\label{L gauge sym metric}
	S_1&=&\int d^4{x} ~
	a_{1}h^{\mu\nu}
	\left[2\partial_{\nu}\partial^{\lambda}h_{\mu\lambda}
	-\Box h_{\mu\nu}
	-2\partial_{\mu}\partial_{\nu}h
	+\eta_{\mu\nu}\Box h\right],
	 \\
	S_I&=&\int d^4{x} ~
	h^{\mu\nu}
	\bigg[
	      \left(a_{I}+b_{I}\right)
	                     \Box^{I-2}\partial_{\mu}\partial_{\nu}\partial^{\lambda}\partial^{\rho}h_{\lambda\rho}
     	-2b_{I}
	          \Box^{I-1}\partial_{\nu}\partial^{\lambda}h_{\mu\lambda}
	\nonumber \\
	&+&b_{I}
	         \Box^{I}h_{\mu\nu}
	-2a_{I}
	         \Box^{I-1}\partial_{\mu}\partial_{\nu}h
	+a_{I} \eta_{\mu\nu}
	                \Box^{I}h
	\bigg].
\end{eqnarray}
It can be seen that if there is no derivative term higher than the second order in the field equation, then this action will degenerate into the Pauli–Fierz action and this is the Einstein-Hilbert action of linearized theory \cite{Michele Maggiore}. Specifically, due to the gauge invariance, we have the relationship (\ref{a(0)is0}). This implies that $S_{0}$, as given by Eq. (\ref{S0 even metric}), is zero. 
Therefore, in the case where the equation only has the highest second-order derivative term, the gauge invariance requires that there is no mass term in the action, thus only describing massless gravitons. It should also be pointed out that when only $a_{1}$ and $a_{2}$ are non-zero, the action returns to the case of $f(R)$ theory. And when only $a_{1}, a_{2}$ and $b_{2}$ are non-zero, the action returns to the case of the  four-dimensional critical gravity \cite{H. Lu1}.

Varying the action (\ref{L gauge sym metric}) with respect to $h^{\mu\nu}$, we can obtain the linearized field equation of general metric theory:
\begin{eqnarray}
	\label{field equation of general metric theory}
	&&2a_{1}\partial_{\nu}\partial^{\lambda}h_{\mu\lambda}
	+2a_{1}\partial_{\mu}\partial^{\lambda}h_{\nu\lambda}
	-2a_{1}\Box h_{\mu\nu}
	\nonumber \\
	&-&2a_{1}\partial_{\mu}\partial_{\nu}h
	-2a_{1}\eta_{\mu\nu}\partial_{\lambda}\partial_{\rho}h^{\lambda\rho}
	+2a_{1}\eta_{\mu\nu}\Box h
	\nonumber \\
	&+&\sum_{I\geq2}^N
	\left[
	   2\left(a_{I}+b_{I}\right)
	          \left(
	            \Box^{I-2}\partial_{\mu}\partial_{\nu}\partial^{\lambda}\partial^{\rho}h_{\lambda\rho}
	          \right)
	  -2b_{I}\Box^{I-1}\partial_{\nu}\partial^{\lambda}h_{\mu\lambda}
	\right.
	\nonumber \\
	&-&2b_{I}\Box^{I-1} \partial_{\mu}\partial^{\lambda}h_{\nu\lambda}
	+2b_{I}\Box^{I}h_{\mu\nu}
	-2a_{I}\Box^{I-1}\partial_{\mu}\partial_{\nu}h
	\nonumber \\
	&-&\left.
	2a_{I}\eta_{\mu\nu}\Box^{I-1}\partial_{\lambda}\partial_{\rho}h^{\lambda\rho}
	+2a_{I}\eta_{\mu\nu}\Box^{I}h
	\right]
	=0.
\end{eqnarray}
We hope that the theory we study includes the structure of general relativity, implying that it should be an extension of general relativity. This requires
\begin{eqnarray}
	\label{a1neq0}
	a_{1} \neq 0.
\end{eqnarray}
In fact, we can always make the normalization $a_{1}$ satisfy $a_1=-\frac{1}{2}$.

\section{Polarization modes of gravitational waves in the general metric theory}
\label{sec: 3}

In this section, we will study the polarizations in the general metric theory. We first introduce a gauge invariant method to simplify Eq. (\ref{field equation of general metric theory}).

We can uniquely decompose the four-dimensional tensor $h_{\mu\nu}$ as follows \cite{James. Bardeen,R. Jackiw,Eanna E Flanagan}:
\begin{eqnarray}
	\label{decompose perturbations}
	h_{00}&=&h_{00}, \nonumber \\
	h_{0i}&=&\partial_{i}\gamma+\beta_{i}, \nonumber \\
	h_{ij}&=&h^{TT}_{ij}+\partial_{i}\epsilon_{j}+\partial_{j}\epsilon_{i}
	+\frac{1}{3}\delta_{ij}H+\left(\partial_{i}\partial_{j}-\frac{1}{3}\delta_{ij}\Delta\right)\zeta,
\end{eqnarray}
where $\beta^{i}$ and $\epsilon^{i}$ are transverse, and $h^{TT}_{ij}$ is transverse and traceless:
\begin{eqnarray}
	&\partial_{i}\beta^{i}=\partial_{i}\epsilon^{i}=0,  \\
	&\delta^{ij}h^{TT}_{ij}=0,~\partial^{i}h^{TT}_{ij}=0.
\end{eqnarray}
Here and below, $\delta_{ij}$ and $\delta^{ij}$ are used to lower and raise the three-dimensional space indices, respectively.

These perturbations can be combined into gauge invariants. Under the gauge transformation (\ref{gauge transformation h}), one can show that gauge invariants include \cite{Eanna E Flanagan} one spatial tensor
\begin{eqnarray}
	\label{tensor gauge invariant}
	h^{TT}_{ij},
\end{eqnarray}
one spatial vector
\begin{eqnarray}
	\label{vector gauge invariant}
	\begin{array}{l}
		\Xi_{i}=\beta_{i}-\partial_{0}\epsilon_{i},
	\end{array}	
\end{eqnarray}
and two spatial scalars
\begin{eqnarray}	
	\begin{array}{l}
		\phi \,= -\frac{1}{2}h_{00}+\partial_{0}\gamma-\frac{1}{2}\partial_{0}\partial_{0}\zeta,  \\
		\Theta = \frac{1}{3}\left(H-\Delta\zeta\right).
	\end{array} \label{scalar gauge invariant}
\end{eqnarray}
The linearized field equation (\ref{field equation of general metric theory}) can be represented by these gauge invariants.

Similar to Eq. (\ref{decompose perturbations}), by decomposing the linearized field equation (\ref{field equation of general metric theory}), one can obtain a series of decoupled scalar, vector, and tensor equations \cite{Weinberg}. These equations simplify the analysis of the polarization modes of gravitational waves.

Now, we review some basic knowledge about the polarization modes of gravitational waves. To detect gravitational waves, we need to detect the relative displacement between free test particles \cite{MTW}. As mentioned in the introduction, the theory we are studying needs to satisfy the condition (4) given in Sec. \ref{sec: intro}, where the action of a free particle is
\begin{eqnarray}
	\label{a free particle}
	S=\int ds.
\end{eqnarray}
Therefore, the relative motion between two test particles satisfies the geodesic deviation equation \cite{MTW}:
\begin{eqnarray}
	\label{equation of geodesic deviation}
	\frac{d^{2}\eta_{i}}{dt^{2}}=-\mathop{R}^{(1)}\!_{i0j0}\eta^{j}.
\end{eqnarray}
Here, $\eta_{i}$ represents the relative displacement of the two test particles,  and the notation ``$(1)$" above $R_{i0j0}$ means that we only take the linear order of the $i0j0$ component of the Riemannian tensor.

We define the polarization modes of gravitational waves based on the relative motion of particles \cite{MTW}. From Eq. (\ref{equation of geodesic deviation}), we can see that as long as we know $R_{i0j0}$, we can obtain the relative motion of the test particles. Therefore, $R_{i0j0}$ can be utilized to define the polarization modes of gravitational waves \cite{Eardley}.

In this paper, we consider plane gravitational waves. For a monochromatic plane gravitational wave, $R_{i0j0}$ can be written as
\begin{eqnarray}
	\label{Ri0j0=AEeikx}
	\mathop{R}^{(1)}\!_{i0j0}=A E_{ij} e^{ikx}.
\end{eqnarray}
Here, $k^{\mu}$ is a four-wavevector, $A$ represents the intensity of the wave, and $E_{ij}$ is a symmetric matrix that contains all polarization information of this plane wave and satisfies
\begin{eqnarray}
	\label{EE=1}
	E_{ij}E^{ij}=1.
\end{eqnarray}
In four-dimensional spacetime, $E_{ij}$ has only six independent components. Therefore, the gravitational waves in the four-dimensional modified gravity theory have up to six independent polarization modes. Any plane gravitational wave can be written as a linear combination of these six polarization modes.

In this paper, without loss of generality, we take the wavevector direction as $+z$ direction. We choose to write the components of $R_{i0j0}$ in the following way to define the six polarization modes of gravitational waves \cite{Eardley}:
\begin{eqnarray}
	\label{P1-P6}
	\mathop{R}^{(1)}\!_{i0j0}=\begin{pmatrix}
		P_{4}+P_{6} & P_{5} & P_{2}\\
		P_{5}       & -P_{4}+P_{6}  & P_{3}\\
		P_{2}       &  P_{3}   &   P_{1}
	\end{pmatrix}.
\end{eqnarray}
Here, $P_{1},\cdots,P_{6}$ correspond to the six independent polarization modes of gravitational waves. We show the six polarization modes in Fig. 1.
\begin{figure*}[htbp]
	\makebox[\textwidth][c]{\includegraphics[width=1.2\textwidth]{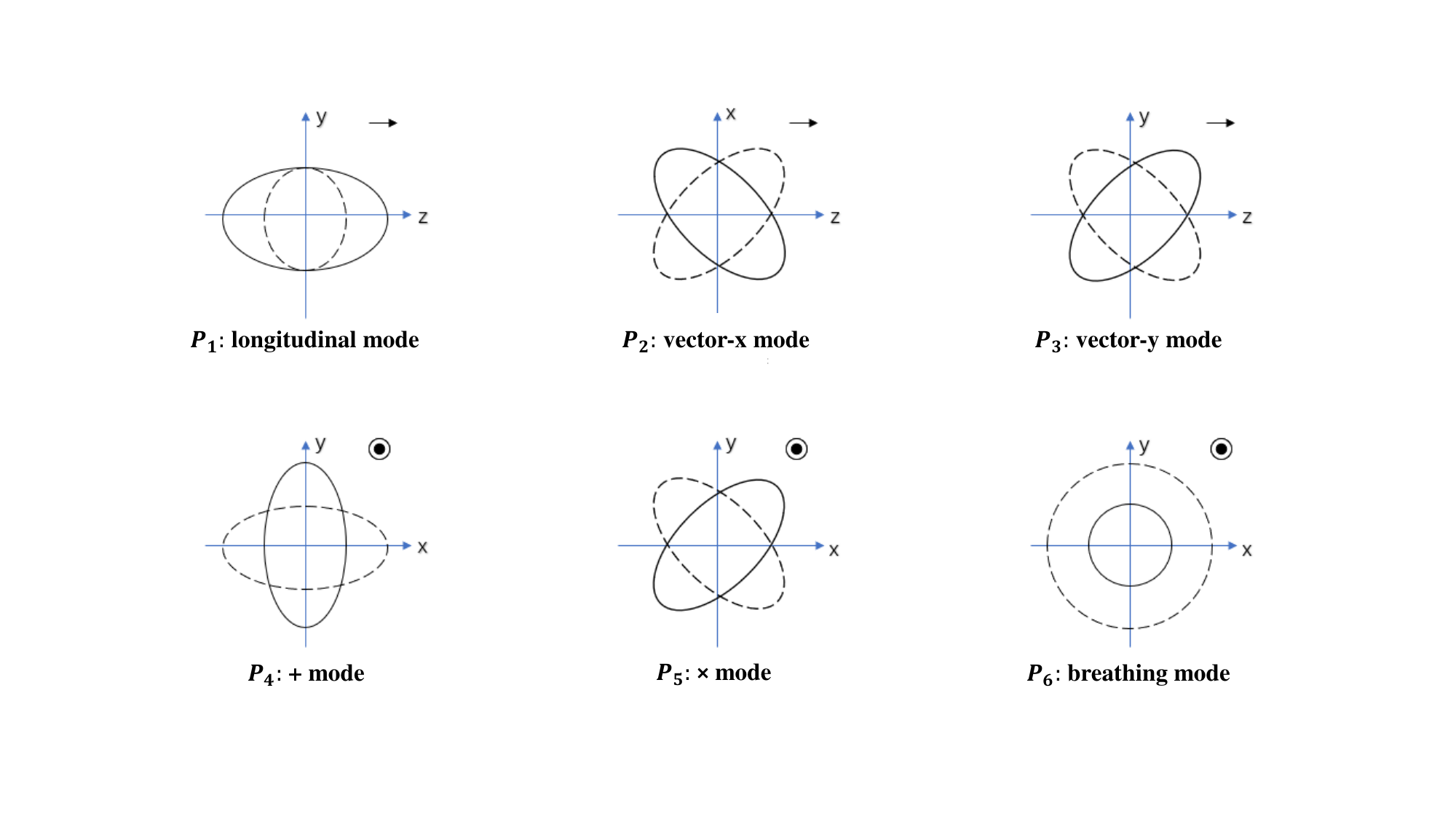}}
	\caption{Six polarization modes of gravitational waves \cite{Eardley}. 
	}
	\label{fig: 1}
\end{figure*}

Using Eqs. (\ref{decompose perturbations}), (\ref{tensor gauge invariant})-(\ref{scalar gauge invariant}), we can write $R_{i0j0}$ as
\begin{eqnarray}
	\label{Ri0j0 gauge invariant}
	\mathop{R}^{(1)}\!_{i0j0}=-\frac{1}{2}\partial_{0}\partial_{0}h^{TT}_{ij}
	+\frac{1}{2}\partial_{0}\partial_{i}\Xi_{j}
	+\frac{1}{2}\partial_{0}\partial_{j}\Xi_{i}
	+\partial_{i}\partial_{j}\phi
	-\frac{1}{2}\delta_{ij}\partial_{0}\partial_{0}\Theta.
\end{eqnarray}
Thus, with Eqs. (\ref{P1-P6}) and (\ref{Ri0j0 gauge invariant}), we can obtain \cite{Y.Dong2}
\begin{eqnarray}
	\label{P1-P6 gauge invariant}
	\begin{array}{l}
		P_{1}=\partial_{3}\partial_{3}\phi-\frac{1}{2}\partial_{0}\partial_{0}\Theta, \quad
		P_{2}=\frac{1}{2}\partial_{0}\partial_{3}\Xi_{1},\\
		P_{3}=\frac{1}{2}\partial_{0}\partial_{3}\Xi_{2},  \quad \quad\quad\quad\,\,
		P_{4}=-\frac{1}{2}\partial_{0}\partial_{0}h^{TT}_{11}, \\
		P_{5}=-\frac{1}{2}\partial_{0}\partial_{0}h^{TT}_{12}, \quad\quad~~~
		P_{6}=-\frac{1}{2}\partial_{0}\partial_{0}\Theta.
	\end{array}
\end{eqnarray}
It can be seen that tensor modes only depend on $h^{TT}_{ij}$ and vector modes only depend on $\Xi_{i}$. For scalar modes, $\phi$ and $\Theta$ contribute to the longitudinal mode, and in addition, $\Theta$ also contributes to the breathing mode.

Now, we will analyze the polarization modes of gravitational waves allowed by the field equation (\ref{field equation of general metric theory}).

\subsection{Tensor modes}
\label{tensor metric}
The equation of the tensor is
\begin{eqnarray}
	\label{tensor mode equation of general metric theory}
	-a_{1}\Box h^{TT}_{ij}
	+\sum_{I\geq2}^N  b_{I} \Box^{I} h^{TT}_{ij}
	=0.
\end{eqnarray}
 Inspired by the analysis in Refs. \cite{H. Lu1,H. Lu2}, we can use the method of solving an equation of degree $N-1$ with one unknown to express the operator $\sum_{I\geq2}^N  b_{I} \Box^{I-1}-a_{1}$ as
 \begin{eqnarray}
 	\label{tensor box operator to 1 n-1 eq}
 	\Lambda_{1}\prod_{k=1}^{M} \left(\Box-{m_{k}}^2\right)^{n_{k}}.
 \end{eqnarray}
Here, $m_{k}$ $(k=1,2,...,M)$ are all roots of the following equation
\begin{eqnarray}
	\label{equation of degree N-1 with one unknown}
	\sum_{I\geq2}^N  b_{I} x^{I-1}-a_{1}=0,
\end{eqnarray}
$n_{k}$ is the multiplicity corresponding to the root $m_{k}$, $\Lambda_{1}\neq0$ is the coefficient of the highest order term in Eq. (\ref{equation of degree N-1 with one unknown}), and $M$ is the total number of different roots.

From Eq. (\ref{tensor box operator to 1 n-1 eq}), Eq. (\ref{tensor mode equation of general metric theory}) can be rewritten as
\begin{eqnarray}
	\label{tensor mode equation of general metric theory simply}
	\Box \prod_{k=1}^{M} \left(\Box-{m_{k}}^2\right)^{n_{k}} h^{TT}_{ij}=0.
\end{eqnarray}
We study the solution with the following plane gravitational wave form:
\begin{eqnarray}
	\label{hijtt=aeikx}
	h^{TT}_{ij}=A_{ij} e^{ikx},
\end{eqnarray}
where $A_{ij}$ is a constant tensor. At this point, due to a separate operator $\Box$ on the left side of Eq. (\ref{tensor mode equation of general metric theory simply}), the general metric theory must have tensor mode gravitational waves propagating at the speed of light.

If each $m_{k}$ is non-zero, then tensor gravitational waves have a total of $M+1$ different masses $\left\{0, m_{1}, m_{2},..., m_{M} \right\}$. Otherwise, tensor gravitational waves have $M$ different masses $\left\{m_{1}, m_{2},..., m_{M} \right\}$. At each mass, tensor gravitational waves have two polarization modes: the $+$ mode and the $\times$ mode. Furthermore, we expect each ${m_{k}}^2$ to be a non-negative real number. This is because taking the imaginary value of ${m_{k}}^2$ would lead to exponential divergence of solution (\ref{hijtt=aeikx}), thereby rendering the theory linearly unstable. And when ${m_{k}}^2$ is less than $0$, the group speed of gravitational waves exceeds the speed of light.

\subsection{Vector modes}
\label{vector metric}
The equations of the two vectors are given by
\begin{eqnarray}
	\label{vector mode equation 1 metric}
	-a_{1} \Delta \Xi_{i}
	+\sum_{I\geq2}^N  b_{I} \Box^{I-1} \Delta \Xi_{i}
	=0, \\
	\label{vector mode equation 2 metric}
	-a_{1} \partial_{0} \Xi_{i}
	+\sum_{I\geq2}^N  b_{I} \Box^{I-1} \partial_{0} \Xi_{i}
	=0.
\end{eqnarray}

We study the solution with the following plane gravitational wave form:
\begin{eqnarray}
	\label{Xii=aeikx}
	\Xi_{i}=A_{i} e^{ikx},
\end{eqnarray}
where $A_{i}$ is a constant vector. In the case of waves we are studying, Eqs. (\ref{vector mode equation 1 metric}) and (\ref{vector mode equation 2 metric}) are equivalent to the following equation:
\begin{eqnarray}
	\label{vector mode equation metric}
	-a_{1} \Xi_{i}
	+\sum_{I\geq2}^N b_{I} \Box^{I-1} \Xi_{i}
	=0.
\end{eqnarray}
Therefore, only Eq. (\ref{vector mode equation metric}) needs to be considered. Using Eq. (\ref{tensor box operator to 1 n-1 eq}), vector equation (\ref{vector mode equation metric}) can be rewritten as
\begin{eqnarray}
	\label{vector mode equation metric simply}
	\prod_{k=1}^{M} \left(\Box-{m_{k}}^2\right)^{n_{k}} \Xi_{i}
	=0.
\end{eqnarray}
It can be seen that in the general metric theory, vector gravitational waves have $M$ different masses $\left\{m_{1}, m_{2},..., m_{M} \right\}$. At each mass, vector gravitational waves have two polarization modes: the vector-x mode and the vector-y mode.

\subsection{Scalar modes}
\label{scalar metric}
We can derive four equations for the four scalars from Eq. (\ref{field equation of general metric theory}):
\begin{eqnarray}
	\label{scalar mode equation 1 metric}
	&-&2a_{1} \Delta \Theta
	+\sum_{I\geq2}^N
	  \left[
	    -2a_{I}\Box^{I-2}\Delta^{2} \phi
	    -2a_{I}\Box^{I-2}\Delta^{2} \Theta
	  \right.
	  \nonumber \\
	&+&\left. 3 a_{I} \Box^{I-2} \partial_{0}^{2} \Delta \Theta
	-2b_{I}\Box^{I-2}\Delta^{2}\phi
	+b_{I}\Box^{I-2}\partial_{0}^{2}\Delta\Theta
	\right]
	=0.
\end{eqnarray}
\begin{eqnarray}
	\label{scalar mode equation 2 metric}
	&-&2a_{1} \partial_{0} \Theta
	+\sum_{I\geq2}^N
	\left[
	-2a_{I}\Box^{I-2}\partial_{0} \Delta \phi
	-2a_{I}\Box^{I-2}\partial_{0} \Delta \Theta
	\right.
	\nonumber \\
	&+&\left. 3 a_{I} \Box^{I-2} \partial_{0}^{3} \Theta
	-2b_{I}\Box^{I-2} \partial_{0} \Delta \phi
	+b_{I}\Box^{I-2}\partial_{0}^{3} \Theta
	\right]
	=0.
\end{eqnarray}
\begin{eqnarray}
	\label{scalar mode equation 3 metric}
	&-&2a_{1} \phi
	-a_{1} \Theta
	+\sum_{I\geq2}^N
	\left[
	   -2a_{I}\Box^{I-2}\Delta\phi
	\right.
	\nonumber \\
	&-&2a_{I}\Box^{I-2}\Delta\Theta
	+3a_{I}\Box^{I-2}\partial_{0}^{2}\Theta
	-2b_{I}\Box^{I-2}\partial_{0}^{2}\phi
	\nonumber \\
	&-&\left.
	b_{I}\Box^{I-2}\Delta\Theta
	+2b_{I}\Box^{I-2}\partial_{0}^{2}\Theta
	\right]
	=0,
\\ \nonumber
\\
	\label{scalar mode equation 4 metric}
	&-&a_{1} \Box \Theta
	+2a_{1} \Delta \phi
	-3a_{1} \partial_{0}^{2}\Theta
	\nonumber \\
	&+&2a_{1} \Delta \Theta
	+\sum_{I\geq2}^N
	\left[
	   b_{I}\Box^{I}\Theta
	   +2a_{I}\Box^{I-1}\Delta\phi
	\right.
	\nonumber \\
	&-&\left.
	3a_{I}\Box^{I-1}\partial_{0}^{2}\Theta
	+2a_{I}\Box^{I-1}\Delta\Theta
	\right]
	=0.
\end{eqnarray}

We study the solution with the following plane gravitational wave form:
\begin{eqnarray}
	\label{Thetaphi=BCeikx}
	\Theta=B e^{ikx}, ~\phi=C e^{ikx},~ k^{\mu}k_{\mu}=-m^2,
\end{eqnarray}
where $B, C$ are constants and $m$ is the mass of the plane gravitational wave. In the case of waves we are studying, there are only two independent equations describing scalar mode gravitational waves obtained from Eqs. (\ref{scalar mode equation 1 metric})-(\ref{scalar mode equation 4 metric}):
\begin{eqnarray}
	\label{scalar mode equation 1 metric simply}
	 &-&a_{1}\left(2\phi-\Theta\right)
	 +\sum_{I\geq2}^N
	 b_{I}\Box^{I-1}\left(2\phi-\Theta\right)
	 =0,
	\\ \label{scalar mode equation 2 metric simply}
	&-&2a_{1}\Theta
	+\sum_{I\geq2}^N
	\left[-\left(3a_{I}+b_{I}\right)\Box^{I-1}\Theta\right]
	=\sum_{I\geq2}^N
	\left(a_{I}+b_{I}\right)\Box^{I-2}\Delta\left(2\phi-\Theta\right).
\end{eqnarray}

Similar to Eq. (\ref{tensor box operator to 1 n-1 eq}), we can rewrite the following operators in Eqs. (\ref{scalar mode equation 1 metric simply}) and (\ref{scalar mode equation 2 metric simply}) as
\begin{eqnarray}
	\label{scalar mode operator simply 1}
	\sum_{I\geq2}^N b_{I}\Box^{I-1}-a_{1}
	&=&
	\Lambda_{1}\prod_{k=1}^{M} \left(\Box-{m_{k}}^2\right)^{n_{k}},
	 \\ \label{scalar mode operator simply 2}
	 \sum_{I\geq2}^N \left[-\left(3a_{I}+b_{I}\right)\Box^{I-1}\right]-2a_{1}
	 &=&
	 \Lambda_{2}\prod_{k=1}^{P} \left(\Box-{\tilde{m}_{k}}^2\right)^{p_{k}},
	 \\ \label{scalar mode operator simply 3}
	 \sum_{I\geq2}^N \left(a_{I}+b_{I}\right)\Box^{I-2}
	 &=&
	 \Lambda_{3}\prod_{k=1}^{Q} \left(\Box-{{m^\prime_{k}}}^2\right)^{q_{k}}.
\end{eqnarray}
Thus, Eqs. (\ref{scalar mode equation 1 metric simply}) and (\ref{scalar mode equation 2 metric simply}) have the form
\begin{eqnarray}
	\label{scalar mode equation 1 metric simply simply}
	&&\prod_{k=1}^{M} \left(\Box-{m_{k}}^2\right)^{n_{k}}
	\left(2\phi-\Theta\right)
	=0,
	\\ \label{scalar mode equation 2 metric simply simply}
	&&\Lambda_{2}\prod_{k=1}^{P} \left(\Box-{\tilde{m}_{k}}^2\right)^{p_{k}} \Theta
	=
	\Lambda_{3} \Delta \prod_{k=1}^{Q} \left(\Box-{{m^\prime_{k}}}^2\right)^{q_{k}}
	\left(2\phi-\Theta\right).
\end{eqnarray}

By substituting the plane wave form (\ref{Thetaphi=BCeikx}) into the above equations, we find that
\begin{eqnarray}
	\label{scalar mode equation 1 metric simply simply plane wave}
	&&\prod_{k=1}^{M} \left(m^2-{m_{k}}^2\right)^{n_{k}}
	\left(2C-B\right)
	=0,
	\\ \label{scalar mode equation 2 metric simply simply plane wave}
	&&\Lambda_{2}\prod_{k=1}^{P} \left(m^2-{\tilde{m}_{k}}^2\right)^{p_{k}} B
	=
	-k_{3}^{2} \Lambda_{3}\prod_{k=1}^{Q} \left(m^2-{{m^\prime_{k}}}^2\right)^{q_{k}}
	\left(2C-B\right).
\end{eqnarray}
Now, the problem is reduced to solve the above two linear equations. We denote the set of elements $m_{k}^{2}$, $\tilde{m}_{k}^{2}$, and ${{m^\prime_{k}}}^2$ as $\left\{m_{k}^{2}\right\}$, $\left\{\tilde{m}_{k}^{2}\right\}$, and $\left\{{{m^\prime_{k}}}^2\right\}$, respectively. The scalar mode gravitational wave at this time can be divided into the following cases for discussion.

\textbf{Case 1.1}: $m^{2} \notin \left\{m_{k}^{2}\right\}, m^{2} \notin \left\{\tilde{m}_{k}^{2}\right\}$. In this case, from Eq. (\ref{scalar mode equation 1 metric simply simply plane wave}), it can be inferred that $2C-B=0$. Furthermore, from Eq. (\ref{scalar mode equation 2 metric simply simply plane wave}), it can be inferred that $B=0$. Therefore, $B=C=0$. In this case, there is no scalar gravitational wave with a mass of $m$.

\textbf{Case 1.2}: $m^{2} \notin \left\{m_{k}^{2}\right\}, m^{2} \in \left\{\tilde{m}_{k}^{2}\right\}$. In this case, we still have $2C-B=0$. In addition, Eq. (\ref{scalar mode equation 2 metric simply simply plane wave}) is naturally satisfied. Therefore, there is a scalar gravitational wave with a mass of $m$ that satisfies the relationship $C=\frac{B}{2}$.

\textbf{Case 2.1}: $m^{2} \in \left\{m_{k}^{2}\right\}, m^{2} \notin \left\{\tilde{m}_{k}^{2}\right\}, m^{2} \notin \left\{{m^\prime_{k}}^{2}\right\}$. In this case, Eq. (\ref{scalar mode equation 1 metric simply simply plane wave}) is naturally satisfied and Eq. (\ref{scalar mode equation 2 metric simply simply plane wave}) reduces to
\begin{eqnarray}
	\label{relationship B and C}
	C=\frac{1}{2}\left(1-\frac{\Lambda_{2}\prod_{k=1}^{P} \left(m^2-{\tilde{m}_{k}}^2\right)^{p_{k}} }{k_{3}^{2} \Lambda_{3}\prod_{k=1}^{Q} \left(m^2-{{m^\prime_{k}}}^2\right)^{q_{k}}}\right)B.
\end{eqnarray}
Therefore, in this case, there is a scalar gravitational wave with a mass of $m$ that satisfies the relationship (\ref{relationship B and C}).

\textbf{Case 2.2}: $m^{2} \in \left\{m_{k}^{2}\right\}, m^{2} \notin \left\{\tilde{m}_{k}^{2}\right\}, m^{2} \in \left\{{m^\prime_{k}}^{2}\right\}$. In this case, the solution is $B=0$ and there is a scalar gravitational wave with mass $m$ and $B=0$.

\textbf{Case 2.3}: $m^{2} \in \left\{m_{k}^{2}\right\}, m^{2} \in \left\{\tilde{m}_{k}^{2}\right\}, m^{2} \notin \left\{{m^\prime_{k}}^{2}\right\}$. The solution for this case is $C=\frac{B}{2}$. Therefore, there is a scalar gravitational wave with mass $m$ and $C=\frac{B}{2}$.

\textbf{Case 2.4}: $m^{2} \in \left\{m_{k}^{2}\right\}, m^{2} \in \left\{\tilde{m}_{k}^{2}\right\}, m^{2} \in \left\{{m^\prime_{k}}^{2}\right\}$. In this case, equations (\ref{scalar mode equation 1 metric simply simply plane wave}) and (\ref{scalar mode equation 2 metric simply simply plane wave}) are naturally satisfied. Therefore, there is a scalar gravitational wave with mass $m$ and independent $B$ and $C$.

As can be seen, as long as $m^{2} \in \left\{m_{k}^{2}\right\} \cup \left\{\tilde{m}_{k}^{2}\right\}$, there exists a non zero solution that satisfies the form of (\ref{Thetaphi=BCeikx}). To avoid superluminal speed and linear instability, we also expect each $\tilde{m}_{k}^{2}$ to be a non-negative real number.

Now, we study the polarization modes of the scalar gravitational wave. Substituting the plane gravitational wave solution (\ref{Thetaphi=BCeikx}) into Eq. (\ref{P1-P6 gauge invariant}), it can be seen that the corresponding amplitudes of the longitudinal mode and the breathing mode are
\begin{eqnarray}
	\label{breathing mode and the longitudinal mode}
	P_{1}&=&\left|\left(m^{2}-\omega^{2}\right)C+\frac{1}{2}\omega^{2}B\right|, \quad
    \nonumber \\
	P_{6}&=&\left|\frac{1}{2}\omega^{2} B\right|.
\end{eqnarray}
Here, $\omega=k^{0}$ is the frequency of the scalar gravitational wave.

For Case 1.2 and Case 2.3, we have $2C-B=0$. The corresponding scalar gravitational wave is a mixture of the longitudinal mode to the breathing mode, and the amplitude ratio $\mathcal{R}$ of the longitudinal mode to the breathing mode is
\begin{eqnarray}
	\label{ratio P1,P6}
	\mathcal{R}=
	\left|\frac{P_1}{P_6}\right|=
      \frac{m^{2}}{\omega^{2}}.
\end{eqnarray}
When $\mathcal{R}=0$, i.e., $m=0$, the scalar gravitational wave degenerates into a breathing mode.

For Case 2.1, we have the relationship (\ref{relationship B and C}). The scalar gravitational wave is a mixture of the longitudinal mode to the breathing mode, and the amplitude ratio $\mathcal{R}$ of the longitudinal mode to the breathing mode is
\begin{eqnarray}
	\label{ratio P1,P6 2}
	\mathcal{R}=
	\left|\frac{P_1}{P_6}\right|=
	\left|
	\frac{m^{2}{\Lambda_{3}\prod_{k=1}^{Q} \left(m^2-{{m^\prime_{k}}}^2\right)^{q_{k}}}
		+{\Lambda_{2}\prod_{k=1}^{P} \left(m^2-{\tilde{m}_{k}}^2\right)^{p_{k}} } }
	{\omega^{2}{\Lambda_{3}\prod_{k=1}^{Q} \left(m^2-{{m^\prime_{k}}}^2\right)^{q_{k}}}}
	\right|.
\end{eqnarray}

For Case 2.2, the constraint is $B=0$. Thus, we have $P_{6}=0$ and $P_{1}=\left|\left(m^{2}-\omega^{2}\right)C\right|$. The scalar gravitational wave is a longitudinal mode.

Finally, for Case 2.4, there is no equation limiting the values of $B$ and $C$. Therefore, the scalar gravitational wave could be a breathing mode, a longitudinal mode, or a mixed mode of the two.

We summarize the results of scalar gravitational waves in Tab. \ref{tab:scalar mode in the general metric theory}:
\begin{center}
	\begin{table}[htbp]
		\resizebox{\textwidth}{!}{
			\begin{tabular}{|c|l|c|c|c|c|}
				\hline
				\hline
				\textbf{Cases} & \qquad\textbf{Conditions} & Breathing mode & Longitudinal mode & Dependent or not & $\mathcal{R}=m^2/{\omega}^2$ \\
				\hline
				case 1.1 &$m^{2} \notin \left\{m_{k}^{2}\right\}, m^{2} \notin \left\{\tilde{m}_{k}^{2}\right\}$& $\times$ & $\times$ & - & - \\
				\hline
				case 1.2 &$m^{2} \notin \left\{m_{k}^{2}\right\}, m^{2} \in \left\{\tilde{m}_{k}^{2}\right\}$& $\checkmark$ & $\checkmark$ & $\checkmark$ & $\checkmark$ \\
				\hline
				case 2.1 &$m^{2} \in \left\{m_{k}^{2}\right\}, m^{2} \notin \left\{\tilde{m}_{k}^{2}\right\}, m^{2} \notin \left\{{m^\prime_{k}}^{2}\right\}$& $\checkmark$ & $\checkmark$ & $\checkmark$ & $\times$ \\
				\hline
				case 2.2 &$m^{2} \in \left\{m_{k}^{2}\right\}, m^{2} \notin \left\{\tilde{m}_{k}^{2}\right\}, m^{2} \in \left\{{m^\prime_{k}}^{2}\right\}$& $\times$ & $\checkmark$ & - & - \\
				\hline
				case 2.3 &$m^{2} \in \left\{m_{k}^{2}\right\}, m^{2} \in \left\{\tilde{m}_{k}^{2}\right\}, m^{2} \notin \left\{{m^\prime_{k}}^{2}\right\}$& $\checkmark$ & $\checkmark$ & $\checkmark$ & $\checkmark$ \\
				\hline
				case 2.4 &$m^{2} \in \left\{m_{k}^{2}\right\}, m^{2} \in \left\{\tilde{m}_{k}^{2}\right\}, m^{2} \in \left\{{m^\prime_{k}}^{2}\right\}$& $\checkmark$ & $\checkmark$ & $\times$ & - \\
				\hline
				\hline
		\end{tabular}  }
		\caption{Scalar mode gravitational waves in various cases in the most general metric theory. In this table, the ``$\checkmark$" and ``$\times$" in the third/fourth column indicate, respectively, the existence and nonexistence of a breathing/longitudinal mode for the considered case. The ``$\checkmark$" and ``$\times$" in the fifth column indicate, respectively, that the breathing mode and longitudinal mode are not independent and are independent. The ``$\checkmark$" and ``$\times$" in the sixth column indicate, respectively, that the amplitude ratio of the longitudinal mode and the breathing mode is and is not $\mathcal{R}=m^2/{\omega}^2$.
}
		\label{tab:scalar mode in the general metric theory}
	\end{table}
\end{center}

\section{Some propositions on polarizations of gravitational waves in the general metric theory}
\label{sec: 4}
In the previous section, we studied the polarizations in the general metric theory. It can be seen that starting from the results in the previous section, we can derive some universal properties of the polarization modes of gravitational waves in the general metric theory. In this section, we will list these universal properties and provide the corresponding proofs.

\textbf{Proposition 1}: If the general metric theory allows vector gravitational waves with a mass of $m$, then there must be tensor gravitational waves with a mass of $m$ in the theory. Correspondingly, if the theory allows tensor gravitational waves with a mass of $m \neq 0$, then there must be vector gravitational waves with a mass of $m$ in the theory.

\textbf{Proof}: The above proposition can be easily observed from Eqs. (\ref{tensor mode equation of general metric theory simply}) and (\ref{vector mode equation metric simply}). According to Eq. (\ref{vector mode equation metric simply}), the theory allows vector gravitational waves with a mass of $m$ to be equivalent to $m \in \left\{m_{k}\right\}$. Therefore, according to Eq. (\ref{tensor mode equation of general metric theory simply}), this theory also allows for tensor gravitational waves with a mass of $m$. This completes the proof of the first sentence of the proposition. The proof in the latter sentence is entirely similar.$\hfill\square$

\textbf{Proposition 2}: If the general metric theory allows tensor gravitational waves propagating only at the speed of light, then there must be no vector gravitational waves in the theory.

\textbf{Proof}: At present, we have not found any evidence that tensor gravitational waves do not propagate at the speed of light. Therefore, it is necessary to study the case where the theory only allows tensor gravitational waves propagating at the speed of light, which shows that the linear perturbation equation describing them will inevitably have the following form:
\begin{eqnarray}
	\label{tensor mode equation metric simply speed of light}
	\Box^{n} h^{TT}_{ij}=0.
\end{eqnarray}
Here, $n \in \left\{1, 2, ..., N\right\}$. Comparing Eq. (\ref{tensor mode equation metric simply speed of light}) with Eq. (\ref{tensor mode equation of general metric theory}), and considering the condition (\ref{a1neq0}), it can be seen that the general metric theory with tensor gravitational waves propagating only at the speed of light requires
\begin{eqnarray}
	\label{bI=0}
	b_{I}=0, ~  I \in \left\{2, 3, ..., N\right\}.
\end{eqnarray}
Thus, Eq. (\ref{vector mode equation metric}) will become $-2a_{1} \Xi_{i}=0$. This requires $\Xi_{i}=0$, and the theory does not allow vector gravitational waves.$\hfill\square$

\textbf{Proposition 3}: If the theory allows tensor gravitational waves propagating only at the speed of light, then for a scalar gravitational wave with mass $m$ and frequency $\omega$, it must be a mixed mode of the longitudinal mode and breathing mode. And the amplitude ratio of the longitudinal mode to the breathing mode must be $m^2/{\omega}^2$ (Therefore, when $m=0$, it degenerates into a breathing mode).

\textbf{Proof}: The general metric theory with tensor gravitational waves propagating only at the speed of light is equivalent to satisfying the condition (\ref{bI=0}), from which we know that Eq. (\ref{scalar mode equation 1 metric simply}) requires $2\phi-\Theta=0$. According to Eq. (\ref{ratio P1,P6}), this requires that for a scalar gravitational wave with mass $m$ and frequency $\omega$, the amplitude ratio of the longitudinal mode to the breathing one is $m^2/{\omega}^2$.$\hfill\square$

\textbf{Proposition 4}: If the amplitude ratio of the longitudinal mode to the breathing one of a scalar gravitational wave with mass $m$ and frequency $\omega$ is not $m^2/{\omega}^2$, then the theory must have a field equation higher than the second derivative and must have tensor and vector gravitational waves with speed less than the speed of light.

\textbf{Proof}: If the amplitude ratio of the longitudinal mode to the breathing one of a scalar gravitational wave with mass $m$ and frequency $\omega$ is not $m^2/{\omega}^2$, then there must exist $J \in \left\{2, 3, ..., N\right\}$, such that $b_{J} \neq 0$. In this case, there must be a $2J$-th order derivative term in Eq. (\ref{tensor mode equation of general metric theory}). Therefore,  the theory must have a field equation higher than the second derivative. In addition, according to the proof of Proposition 2, $b_{J} \neq 0$ means that the theory allows a tensor gravitational wave with a mass $m \neq 0$. By Proposition 1, the theory also allows for massive vector gravitational waves.
$\hfill\square$

\section{Second-order action of the general scalar-tensor theory}
\label{sec: 5}
We have completed the study of the polarizations in the most general metric theory. In this section, we will turn to construct the most general second-order action of the general scalar-tensor theory.

For the general scalar-tensor gravity, in addition to the metric $g_{\mu\nu}$, there is an additional scalar field $\Psi$ used to describe gravity. We assume that the general scalar-tensor theory has a flat spacetime background solution. Thus, we can consider a perturbation on the background:
\begin{eqnarray}
	\label{scalar-tensor background perturbation}
	g_{\mu\nu}=\eta_{\mu\nu}+h_{\mu\nu},~\Psi=\Psi_{0}+\psi.
\end{eqnarray}
Here, $\eta_{\mu\nu}$ is the Minkowski background metric, $\Psi_{0}$ is the constant background scalar field, and $h_{\mu\nu}$ and $\psi$ are perturbations of the background metric and the background scalar field, respectively. In this way, the second-order action of the general scalar-tensor theory will be a functional of $h_{\mu\nu}$ and $\psi$.

Just like the general metric theory, when applying the conditions given in Sec. {\ref{sec: intro}} to the second-order action of the most general four-dimensional scalar-tensor theory, we have the following requirements:
\begin{itemize}
	\item [(1)] Each term in the action is a second order term of $h_{\mu\nu}$ and $\psi$, and it is a combination of $\eta_{\mu\nu}$, $\partial_{\mu}$, $h_{\mu\nu}$, $\psi$, and the coupling parameters of the theory.
	\item [(2)] The action is invariant under the gauge transformation $x^{\mu} \rightarrow x^{\mu}+\xi^{\mu}(x)$.
\end{itemize}
Note that the action is also independent of $E_{\mu\nu\lambda\rho}$, and each term in the action has an even number of derivatives.

Here, we consider the most general second-order action that can
derive the field equations with terms up to $2N$-th derivative terms. This action consists of three parts. The first part is constructed from $h$, the second part is from $\psi$, and the third part is from both $h$ and $\psi$.

The most general form of the first part is exactly the same as that of the most general metric theory, that is, it has the same form as Eq. (\ref{S=Sigma SI metric}). The most general forms of the second and third parts are
\begin{eqnarray}
	\label{The most general form psi psi}
	\int d^4{x}
	\sum_{I=0}^N
	c_{1}^{(I)}\left(\Box^{I}\psi\right)\psi,
\end{eqnarray}
and
\begin{eqnarray}
	\label{The most general form h psi}
	\int d^4{x} \Bigg[
	\sum_{I=1}^N
	\left(
	d_{1}^{(I)}\Box^{I-1}\partial_{\mu}\partial_{\nu}h^{\mu\nu}
	+d_{2}^{(I)}\Box^{I}h
    \right)\psi
	+d_{2}^{(0)}h\psi
	\Bigg],
\end{eqnarray}
respectively.
For convenience, we also define,
\begin{eqnarray}
	\label{d(0)1=0}
	d_{1}^{(0)}=0.
\end{eqnarray}

Under the transformation (\ref{gauge x->x+xi}), $\psi$ is invariant. Therefore, the action (\ref{The most general form psi psi}) is naturally gauge invariant. The terms in the action (\ref{The most general form h psi}) transform as follows:
\begin{eqnarray}
	\label{d1 change}
   d_{1}^{(I)}\left(\Box^{I-1}\partial_{\mu}\partial_{\nu}h^{\mu\nu}\right)\psi                        &\rightarrow &
   d_{1}^{(I)}\left(\Box^{I-1}\partial_{\mu}\partial_{\nu}h^{\mu\nu}\right)\psi
   \nonumber \\
   &-&2d_{1}^{(I)}\left(\Box^{I}\partial_{\mu}\xi^{\mu}\right)\psi,
\nonumber \\
	\label{d2 change}
	d_{2}^{(I)}\left(\Box^{I}h\right)\psi
	 &\rightarrow &
	d_{2}^{(I)}\left(\Box^{I}h\right)\psi
	\nonumber \\
	&-&2d_{2}^{(I)}\left(\Box^{I}\partial_{\mu}\xi^{\mu}\right)\psi.
\end{eqnarray}
It can be seen that the condition of gauge invariance for the second-order action of the general scalar-tensor theory reduces to
\begin{eqnarray}
	\label{d1+d2=0}
	d_{1}^{(I)}+d_{2}^{(I)}=0.
\end{eqnarray}

We redefine $c^{(I)}_{1}$ as $c_{I}$ and $d^{(I)}_{1}$ as $d_{I}$. In this way, considering the gauge symmetry, the most general second-order action that can derive the field equations with up to $2N$-th derivative terms can be written as
\begin{eqnarray}
	S=
	S_{1}+\sum_{I\geq2}^N S_{I},  \label{S gauge sym metric of scalar-tensor}
\end{eqnarray}
where
\begin{eqnarray}
	S_1&=&\int d^4{x}~a_{1}h^{\mu\nu}
	\Big[
	2\partial_{\nu}\partial^{\lambda}h_{\mu\lambda}
	-\Box h_{\mu\nu}
	-2\partial_{\mu}\partial_{\nu}h
	+\eta_{\mu\nu}\Box h
	\Big]
	\nonumber \\
	&+&\int d^4{x}~\psi
	\Big[
	c_{0}\psi
	+c_{1} \Box \psi
	+d_{1} \partial_{\mu}\partial_{\nu}h^{\mu\nu}
	-d_{1} \Box h
	\Big],
	\label{S1 gauge sym metric of scalar-tensor} \\	
	S_{I}
	&=&\int d^4{x}~h^{\mu\nu}
	\Big[
	\left(a_{I}+b_{I}\right)
	\Box^{I-2}\partial_{\mu}\partial_{\nu}\partial^{\lambda}\partial^{\rho}h_{\lambda\rho}
	-2b_{I}
	\Box^{I-1}\partial_{\nu}\partial^{\lambda}h_{\mu\lambda}
	\nonumber \\
	&+&b_{I}
	\Box^{I}h_{\mu\nu}
	-2a_{I}
	\Box^{I-1}\partial_{\mu}\partial_{\nu}h
	+a_{I} \eta_{\mu\nu}
	\Box^{I}h
	\Big]
	\nonumber \\
	&+&\int d^4{x}~\psi
	\Big[
	c_{I} \Box^{I}\psi
	+d_{I} \Box^{I-1}\partial_{\mu}\partial_{\nu}h^{\mu\nu}
	-d_{I} \Box^{I}h
	\Big].
\end{eqnarray}
It should be pointed out that only when  $a_{1}$, $c_{0}$, $c_{1}$ and $d_{1}$ are not zero, the action (\ref{S1 gauge sym metric of scalar-tensor}) is just the case of Horndeski theory.

Varying the action (\ref{S gauge sym metric of scalar-tensor}) with respect to $h^{\mu\nu}$, we can obtain
\begin{eqnarray}
	\label{metric field equation of general scalar-tensor theory}
	&&2a_{1}\partial_{\nu}\partial^{\lambda}h_{\mu\lambda}
	+2a_{1}\partial_{\mu}\partial^{\lambda}h_{\nu\lambda}
	-2a_{1}\Box h_{\mu\nu}
	-2a_{1}\partial_{\mu}\partial_{\nu}h
	\nonumber \\
	&-&2a_{1}\eta_{\mu\nu}\partial_{\lambda}\partial_{\rho}h^{\lambda\rho}
	+2a_{1}\eta_{\mu\nu}\Box h
	+d_{1}\partial_{\mu}\partial_{\nu}\psi
	-d_{1}\eta_{\mu\nu}\Box \psi
	\nonumber \\
	&+&\sum_{I\geq2}^N
	\Bigg[
	2\left(a_{I}+b_{I}\right)
	\left(
	\Box^{I-2}\partial_{\mu}\partial_{\nu}\partial^{\lambda}\partial^{\rho}h_{\lambda\rho}
	\right)
	-2b_{I}\Box^{I-1}\partial_{\nu}\partial^{\lambda}h_{\mu\lambda}
	\nonumber \\
	&-&2b_{I}\Box^{I-1} \partial_{\mu}\partial^{\lambda}h_{\nu\lambda}
	+2b_{I}\Box^{I}h_{\mu\nu}
	-2a_{I}\Box^{I-1}\partial_{\mu}\partial_{\nu}h
	\nonumber \\
	&-&
	2a_{I}\eta_{\mu\nu}\Box^{I-1}\partial_{\lambda}\partial_{\rho}h^{\lambda\rho}
	+2a_{I}\eta_{\mu\nu}\Box^{I}h
	\nonumber \\
	&+&
	d_{I}\Box^{I-1} \partial_{\mu}\partial_{\nu}\psi
	-d_{I}\eta_{\mu\nu}\Box^{I}\psi
	\Bigg]
	=0.
\end{eqnarray}
For the same reason as the general metric theory, the condition (\ref{a1neq0}) is still required here. And varying the action (\ref{S gauge sym metric of scalar-tensor}) with respect to $\psi$, we can obtain
\begin{eqnarray}
	\label{scalar field equation of general scalar-tensor theory}
&&
2c_{0}\psi
+2c_{1}\Box \psi
+d_{1} \partial_{\mu}\partial_{\nu}h^{\mu\nu}
-d_{1}\Box h
\nonumber \\
&+&\sum_{I\geq2}^N
\left[
2c_{I}\Box^{I}\psi
+d_{I}\Box^{I-1}\partial_{\mu}\partial_{\nu}h^{\mu\nu}
-d_{I}\Box^{I} h
\right]
=0.
\end{eqnarray}

\section{Polarization modes of gravitational waves in the most general scalar-tensor theory}
\label{sec: 6}
In this section, we will analyze the polarizations in the most general scalar-tensor theory. Similar to the most general metric theory, we will use the gauge invariant method.

For the most general scalar-tensor theory, it is easy to see from Eqs. (\ref{metric field equation of general scalar-tensor theory}) and (\ref{scalar field equation of general scalar-tensor theory}) that the additional scalar perturbation $\psi$ does not contribute to the decoupled tensor and vector equations. So, the equations describing tensor and vector gravitational waves are still Eqs. (\ref{tensor mode equation of general metric theory}) and (\ref{vector mode equation metric}), respectively. Therefore, the speed and polarization properties of tensor and vector modes in the most general scalar-tensor theory are identical to those in the most general metric theory. The results for tensor and vector modes in Sec. \ref{sec: 4} are still fully valid.

Now, we only need to study the scalar modes of gravitational waves. We can derive equations of the five scalars from Eqs. (\ref{metric field equation of general scalar-tensor theory}) and (\ref{scalar field equation of general scalar-tensor theory}):
\begin{eqnarray}
	\label{scalar mode equation 1 scalar-tensor}
	&-&4a_{1} \Delta \Theta
	+d_{1} \Delta\psi
	+\sum_{I\geq2}^N
	\left[
	-4a_{I}\Box^{I-2}\Delta^{2} \phi
	-4a_{I}\Box^{I-2}\Delta^{2} \Theta
	\right.
	\nonumber \\
	&+&\left. 6 a_{I} \Box^{I-2} \partial_{0}^{2} \Delta \Theta
	-4b_{I}\Box^{I-2}\Delta^{2}\phi
	+2b_{I}\Box^{I-2}\partial_{0}^{2}\Delta\Theta
	+d_{I}\Box^{I-1}\Delta\psi
	\right]
	=0.
\end{eqnarray}
\begin{eqnarray}
	\label{scalar mode equation 2 scalar-tensor}
	&-&4a_{1} \partial_{0} \Theta
	+d_{1}\partial_{0}\psi
	+\sum_{I\geq2}^N
	\left[
	-4a_{I}\Box^{I-2}\partial_{0} \Delta \phi
	-4a_{I}\Box^{I-2}\partial_{0} \Delta \Theta
	\right.
	\nonumber \\
	&+&\left. 6 a_{I} \Box^{I-2} \partial_{0}^{3} \Theta
	-4b_{I}\Box^{I-2} \partial_{0} \Delta \phi
	+2b_{I}\Box^{I-2}\partial_{0}^{3} \Theta
	+d_{I} \Box^{I-1} \partial_{0}\psi
	\right]
	=0.
\end{eqnarray}
\begin{eqnarray}
	\label{scalar mode equation 3 scalar-tensor}
	&-&4a_{1} \phi
	-2a_{1} \Theta
	+d_{1}\psi
	+\sum_{I\geq2}^N
	\left[
	-4a_{I}\Box^{I-2}\Delta\phi
	\right.
	\nonumber \\
	&-&4a_{I}\Box^{I-2}\Delta\Theta
	+6a_{I}\Box^{I-2}\partial_{0}^{2}\Theta
	-4b_{I}\Box^{I-2}\partial_{0}^{2}\phi
	\nonumber \\
	&-&\left.
	2b_{I}\Box^{I-2}\Delta\Theta
	+4b_{I}\Box^{I-2}\partial_{0}^{2}\Theta
	+d_{I}\Box^{I-1}\psi
	\right]
	=0,
	\\ \nonumber
	\\
	\label{scalar mode equation 4 scalar-tensor}
	&-&2a_{1} \Box \Theta
	+4a_{1} \Delta \phi
	-6a_{1} \partial_{0}^{2}\Theta
	+4a_{1} \Delta \Theta
	\nonumber \\
	&-&d_{1} \Box \psi
	+\sum_{I\geq2}^N
	\left[
	2b_{I}\Box^{I}\Theta
	+4a_{I}\Box^{I-1}\Delta\phi
	\right.
	\nonumber \\
	&-&\left.
	6a_{I}\Box^{I-1}\partial_{0}^{2}\Theta
	+4a_{I}\Box^{I-1}\Delta\Theta
	-d_{I}\Box^{I}\psi
	\right]
	=0.
\end{eqnarray}
\begin{eqnarray}
	\label{scalar mode equation 5 scalar-tensor}
	&&2c_{0}\psi
	+2c_{1}\Box\psi
	+d_{1}\left(-2\Delta\phi+\Delta\Theta-3\Box\Theta\right)
	\nonumber \\
	&+&\sum_{I\geq2}^N
	\left[
	2c_{I}\Box^{I}\psi
	+d_{I}\Box^{I-1}
	\left(-2\Delta\phi+\Delta\Theta-3\Box\Theta\right)
	\right]
	=0.
\end{eqnarray}

We study the solution with the following plane gravitational wave form:
\begin{eqnarray}
	\label{Thetaphipsi=BCDeikx}
	\Theta=B e^{ikx}, ~\phi=C e^{ikx},~ \psi=D e^{ikx}, ~ k^{\mu}k_{\mu}=-m^2,
\end{eqnarray}
where $B, C, D$ are constants and $m$ is the mass of the plane gravitational wave. In the case of the waves we are studying, there are only three independent equations describing scalar mode gravitational waves, obtained from Eqs. (\ref{scalar mode equation 1 scalar-tensor})-(\ref{scalar mode equation 5 scalar-tensor}):
\begin{eqnarray}
	\label{scalar mode equation 1 scalar-tensor simply}
	&-&a_{1}\left(2\phi-\Theta\right)
	+\sum_{I\geq2}^N
	b_{I}\Box^{I-1}\left(2\phi-\Theta\right)
	=0,
\end{eqnarray}
\begin{eqnarray}
    \label{scalar mode equation 2 scalar-tensor simply}
	&-&4a_{1}\Theta
	+\sum_{I\geq2}^N
	\left[-\left(6a_{I}+2b_{I}\right)\Box^{I-1}\Theta\right]
	\nonumber \\
	&-&\sum_{I\geq2}^N
	\left(2a_{I}+2b_{I}\right)\Box^{I-2}\Delta\left(2\phi-\Theta\right)
	\nonumber \\
	&+&d_{1}\psi
	+\sum_{I\geq2}^N
	d_{I}\Box^{I-1}\psi
	=0,
\end{eqnarray}
\begin{eqnarray}
	\label{scalar mode equation 3 scalar-tensor simply}
	&&2c_{0}\psi
	+2c_{1}\Box\psi
	+\sum_{I\geq2}^N
	2c_{I}\Box^{I}\psi
	\nonumber \\
	&-&d_{1}\Delta\left(2\phi-\Theta\right)
	-\sum_{I\geq2}^N
	d_{I}\Box^{I-1}\Delta
	\left(2\phi-\Theta\right)
	\nonumber \\
	&-&3d_{1}\Box\Theta
	-\sum_{I\geq2}^N
	3d_{I}\Box^{I}\Theta
	=0.
\end{eqnarray}

We can rewrite the following operators in Eqs. (\ref{scalar mode equation 1 scalar-tensor simply}), (\ref{scalar mode equation 2 scalar-tensor simply}) and (\ref{scalar mode equation 3 scalar-tensor simply}) as
\begin{eqnarray}
	\label{scalar mode operator simply 1 scalar-tensor}
	\sum_{I\geq2}^N b_{I}\Box^{I-1}-a_{1}
	&=&
	\Lambda_{1}\prod_{k=1}^{M} \left(\Box-{m_{k}}^2\right)^{n_{k}},
	\\ \label{scalar mode operator simply 2 scalar-tensor}
	\sum_{I\geq2}^N \left[-\left(3a_{I}+b_{I}\right)\Box^{I-1}\right]-2a_{1}
	&=&
	\Lambda_{2}\prod_{k=1}^{P} \left(\Box-{\tilde{m}_{k}}^2\right)^{p_{k}},
	\\ \label{scalar mode operator simply 3 scalar-tensor}
	\sum_{I\geq2}^N \left(a_{I}+b_{I}\right)\Box^{I-2}
	&=&
	\Lambda_{3}\prod_{k=1}^{Q} \left(\Box-{{m^\prime_{k}}}^2\right)^{q_{k}},
  \\ \label{scalar mode operator simply 4 scalar-tensor}
	\sum_{I\geq2}^N d_{I}\Box^{I-1}+d_{1}
	&=&
	\Lambda_{4}\prod_{k=1}^{R} \left(\Box-{{\hat{m}_{k}}}^2\right)^{r_{k}},
	\\ \label{scalar mode operator simply 5 scalar-tensor}
	\sum_{I\geq2}^N 2c_{I}\Box^{I}+2c_{1}\Box+2c_{0}
	&=&
	\Lambda_{5}\prod_{k=1}^{S} \left(\Box-{{\overline{m}_{k}}}^2\right)^{s_{k}}.
\end{eqnarray}
Thus, Eqs. (\ref{scalar mode equation 1 scalar-tensor simply}), (\ref{scalar mode equation 2 scalar-tensor simply}) and (\ref{scalar mode equation 3 scalar-tensor simply}) take the form
\begin{eqnarray}
	\label{scalar mode equation 1 scalar-tensor simply simply}
	\prod_{k=1}^{M} \left(\Box-{m_{k}}^2\right)^{n_{k}}\left(2\phi-\Theta\right)
	\!\!&=&\!\!0,
\\
	\label{scalar mode equation 2 scalar-tensor simply simply}
	2\Lambda_{2}\prod_{k=1}^{P} \left(\Box-{\tilde{m}_{k}}^2\right)^{p_{k}}\Theta
	+\Lambda_{4}\prod_{k=1}^{R} \left(\Box-{{\hat{m}_{k}}}^2\right)^{r_{k}}\psi
	\!\!&=&\!\!
	2\Lambda_{3}\prod_{k=1}^{Q} \left(\Box-{{m^\prime_{k}}}^2\right)^{q_{k}} \!\! \Delta\left(2\phi-\Theta\right)\!\!,
\\
	\label{scalar mode equation 3 scalar-tensor simply simply}
	\Lambda_{5}\prod_{k=1}^{S} \left(\Box-{{\overline{m}_{k}}}^2\right)^{s_{k}}\psi
	-3\Lambda_{4}\prod_{k=1}^{R} \left(\Box-{{\hat{m}_{k}}}^2\right)^{r_{k}}\Box\Theta
	\!\!&=&\!\!
	\Lambda_{4}\prod_{k=1}^{R} \left(\Box-{{\hat{m}_{k}}}^2\right)^{r_{k}}\Delta
	\left(2\phi-\Theta\right).
\end{eqnarray}

By substituting the plane wave form (\ref{Thetaphipsi=BCDeikx}) into the equations above, we find that
\begin{eqnarray}
	\label{scalar mode equation 1 scalar-tensor simply simply simply}
	\prod_{k=1}^{M} \left(m^{2}-{m_{k}}^2\right)^{n_{k}}\left(2C-B\right)
	\!\!&=&\!\!
	0,
	\\
	2\Lambda_{2}\prod_{k=1}^{P} \left(m^{2}-{\tilde{m}_{k}}^2\right)^{p_{k}}B
	+\Lambda_{4}\prod_{k=1}^{R} \left(m^{2}-{{\hat{m}_{k}}}^2\right)^{r_{k}}D
	\!\!&=&\!\!
	-2k_{3}^{2}\Lambda_{3}\prod_{k=1}^{Q}\!\! \left(m^{2}-{{m^\prime_{k}}}^2\right)^{q_{k}}\!\!\left(2C-B\right)\!\!,
	\nonumber\\
		\label{scalar mode equation 2 scalar-tensor simply simply simply}
	\\
	\Lambda_{5}\prod_{k=1}^{S} \left(m^{2}-{{\overline{m}_{k}}}^2\right)^{s_{k}}D
	-3m^{2}\Lambda_{4}\prod_{k=1}^{R} \left(m^{2}-{{\hat{m}_{k}}}^2\right)^{r_{k}}B
	\!\!&=&\!\!
	-k_{3}^{2} \Lambda_{4}\prod_{k=1}^{R} \left(m^{2}-{{\hat{m}_{k}}}^2\right)^{r_{k}}
	\left(2C-B\right).
	\nonumber \\
	\label{scalar mode equation 3 scalar-tensor simply simply simply}
\end{eqnarray}
It is useful to write Eqs. (\ref{scalar mode equation 2 scalar-tensor simply simply simply}) and (\ref{scalar mode equation 3 scalar-tensor simply simply simply}) in matrix form:
\begin{eqnarray}
	\label{scalar mode equation 2 and 3 in matrix form}
	\mathcal{A}X=\mathcal{B},
\end{eqnarray}
where
\begin{eqnarray}
	\label{scalar mode equation 2 and 3 in matrix form AXB}
	\mathcal{A}
	&=&
	\begin{pmatrix}
		2\Lambda_{2}\prod_{k=1}^{P} \left(m^{2}-{\tilde{m}_{k}}^2\right)^{p_{k}}        ~ & ~
		\Lambda_{4}\prod_{k=1}^{R} \left(m^{2}-{{\hat{m}_{k}}}^2\right)^{r_{k}} \\
		-3m^{2}\Lambda_{4}\prod_{k=1}^{R} \left(m^{2}-{{\hat{m}_{k}}}^2\right)^{r_{k}}   ~&  ~
		\Lambda_{5}\prod_{k=1}^{S} \left(m^{2}-{{\overline{m}_{k}}}^2\right)^{s_{k}}
	\end{pmatrix},
	\nonumber \\
	\nonumber \\
	X
	&=&
	\begin{pmatrix}
		B\\D
	\end{pmatrix},~
	\mathcal{B}
	=-k_{3}^{2}\left(2C-B\right)
	\begin{pmatrix}
		2\Lambda_{3}\prod_{k=1}^{Q} \left(m^{2}-{{m^\prime_{k}}}^2\right)^{q_{k}}
		\\
		 \Lambda_{4}\prod_{k=1}^{R} \left(m^{2}-{{\hat{m}_{k}}}^2\right)^{r_{k}}
	\end{pmatrix}.
\end{eqnarray}

We still denote the set of elements $m_{k}^{2}$ as $\left\{m_{k}^{2}\right\}$, the set of elements $\tilde{m}_{k}^{2}$ as $\left\{\tilde{m}_{k}^{2}\right\}$, etc. The scalar mode gravitational waves can be divided into the following cases for discussion.

\begin{center}
\textbf{Case 1}: $m^{2} \notin \left\{m_{k}^{2}\right\}$
\end{center}

In this case, Eq. (\ref{scalar mode equation 1 scalar-tensor simply simply simply}) requires $2C-B=0$ and Eq. (\ref{scalar mode equation 2 and 3 in matrix form}) becomes
\begin{eqnarray}
	\label{scalar mode equation 2 and 3 in matrix form case A}
	\mathcal{A}X=0.
\end{eqnarray}
This can be further divided into more detailed cases for discussion.

\textbf{Case 1.1}: $m^{2} \notin \left\{m_{k}^{2}\right\},\det(\mathcal{A}) \neq 0$. In this case, the coefficient matrix $\mathcal{A}$ of the linear equation system  (\ref{scalar mode equation 2 and 3 in matrix form case A}) is invertible. Therefore, we obtain $X=0$, i.e. $B=D=0$. Furthermore, due to $2C-B=0$, there is $C=0$. The theory does not allow scalar gravitational waves.

\textbf{Case 1.2}: $m^{2} \notin \left\{m_{k}^{2}\right\},\det(\mathcal{A})=0, m^{2} \in \left\{\tilde{m}_{k}^{2}\right\}, m^{2} \in \left\{\hat{m}_{k}^{2}\right\}$. Equations (\ref{scalar mode equation 2 scalar-tensor simply simply simply}) and (\ref{scalar mode equation 3 scalar-tensor simply simply simply}) are not independent. Therefore, to study scalar gravitational waves, we only need to consider Eq. (\ref{scalar mode equation 2 scalar-tensor simply simply simply}). In this case, the equation can be written as
\begin{eqnarray}
	\label{scalar mode equation 2 scalar-tensor simply simply simply simply}
	2\Lambda_{2}\prod_{k=1}^{P} \left(m^{2}-{\tilde{m}_{k}}^2\right)^{p_{k}}B
	+\Lambda_{4}\prod_{k=1}^{R} \left(m^{2}-{{\hat{m}_{k}}}^2\right)^{r_{k}}D
	\!\!&=&\!\!
	0.
\end{eqnarray}
It can be seen that when $m^{2} \in \left\{\tilde{m}_{k}^{2}\right\}$ and $m^{2} \in \left\{\hat{m}_{k}^{2}\right\}$, Eq. (\ref{scalar mode equation 2 scalar-tensor simply simply simply simply}) is naturally satisfied. So, $B$ and $D$ can take any value. Therefore, there is a scalar gravitational wave with mass $m$ and $C=\frac{B}{2}$.

\textbf{Case 1.3}: $m^{2} \notin \left\{m_{k}^{2}\right\},\det(\mathcal{A})=0, m^{2} \in \left\{\tilde{m}_{k}^{2}\right\}, m^{2} \notin \left\{\hat{m}_{k}^{2}\right\}$. In this case, Eq. (\ref{scalar mode equation 2 scalar-tensor simply simply simply simply}) requires $D=0$. There is a scalar gravitational wave with mass $m$ and $C=\frac{B}{2}$.

\textbf{Case 1.4}: $m^{2} \notin \left\{m_{k}^{2}\right\},\det(\mathcal{A})=0, m^{2} \notin \left\{\tilde{m}_{k}^{2}\right\}, m^{2} \in \left\{\hat{m}_{k}^{2}\right\}$. In this case, Eq. (\ref{scalar mode equation 2 scalar-tensor simply simply simply simply}) requires $B=0$. Furthermore, due to $2C-B=0$, there is $C=0$. The theory does not allow scalar gravitational waves.

\textbf{Case 1.5}: $m^{2} \notin \left\{m_{k}^{2}\right\},\det(\mathcal{A})=0, m^{2} \notin \left\{\tilde{m}_{k}^{2}\right\}, m^{2} \notin \left\{\hat{m}_{k}^{2}\right\}$. In this case, Eq. (\ref{scalar mode equation 2 scalar-tensor simply simply simply simply}) requires
\begin{eqnarray}
	\label{relationship of B D in case 1.5}
	B=-\frac{\Lambda_{4}\prod_{k=1}^{R} \left(m^{2}-{{\hat{m}_{k}}}^2\right)^{r_{k}}}{2\Lambda_{2}\prod_{k=1}^{P} \left(m^{2}-{\tilde{m}_{k}}^2\right)^{p_{k}}}D.
\end{eqnarray}
There is a scalar gravitational wave with mass $m$ and $C=\frac{B}{2}$.

\begin{center}
	\textbf{Case 2}: $m^{2} \in \left\{m_{k}^{2}\right\},\det(\mathcal{A}) \neq 0$
\end{center}

In this case, Eq. (\ref{scalar mode equation 1 scalar-tensor simply simply simply}) is naturally satisfied. For Eq. (\ref{scalar mode equation 2 and 3 in matrix form}), the coefficient matrix $\mathcal{A}$ is invertible. Therefore, we have
\begin{eqnarray}
	\label{solution of scalar mode equation 2 and 3 in matrix form}
	X=\mathcal{A}^{-1} \mathcal{B}.
\end{eqnarray}
Here, $\mathcal{A}^{-1}$ is the inverse of $\mathcal{A}$ and can be expressed as:
\begin{eqnarray}
	\label{scalar mode equation 2 and 3 in matrix form AXB}
	\mathcal{A}^{-1}
	&=&
	\frac{1}{det(\mathcal{A})}
	\begin{pmatrix}
		\Lambda_{5}\prod_{k=1}^{S} \left(m^{2}-{{\overline{m}_{k}}}^2\right)^{s_{k}} ~&~
		-\Lambda_{4}\prod_{k=1}^{R} \left(m^{2}-{{\hat{m}_{k}}}^2\right)^{r_{k}}\\
		3m^{2}\Lambda_{4}\prod_{k=1}^{R} \left(m^{2}-{{\hat{m}_{k}}}^2\right)^{r_{k}}  ~&~
		2\Lambda_{2}\prod_{k=1}^{P} \left(m^{2}-{\tilde{m}_{k}}^2\right)^{p_{k}}
	\end{pmatrix}.
\end{eqnarray}
It can be seen that the relationship between $B$ and $C$ satisfies
\begin{eqnarray}
	\label{relationship between B and C case 2}
	B=
	-\gamma k_{3}^{2}\left(2C-B\right),
\end{eqnarray}
where
\begin{eqnarray}
	\label{gamma}
{\det(\mathcal{A})}\gamma&=&
2\Lambda_{3}\Lambda_{5}\prod_{k=1}^{Q} \left(m^{2}-{{m^\prime_{k}}}^2\right)^{q_{k}}
\prod_{k=1}^{S} \left(m^{2}-{{\overline{m}_{k}}}^2\right)^{s_{k}}
\nonumber \\
&-&
\Lambda_{4}^2\prod_{k=1}^{R} \left(m^{2}-{{\hat{m}_{k}}}^2\right)^{r_{k}}
\prod_{k=1}^{R} \left(m^{2}-{{\hat{m}_{k}}}^2\right)^{r_{k}}.
\end{eqnarray}

We can further divide it into more detailed cases for discussion.

\textbf{Case 2.1}: $m^{2} \in \left\{m_{k}^{2}\right\},\det(\mathcal{A}) \neq 0, \gamma=0$. In this case, from Eq. (\ref{relationship between B and C case 2}), we obtain $B=0$. There is a scalar gravitational wave with mass $m$ and $B=0$.

\textbf{Case 2.2}: $m^{2} \in \left\{m_{k}^{2}\right\},\det(\mathcal{A}) \neq 0, \gamma \neq 0$. In this case, from Eq. (\ref{relationship between B and C case 2}), we obtain
\begin{eqnarray}
	\label{relationship between B and C case 2.2}
	C=
	\frac{\gamma k_{3}^{2}-1}{2\gamma k_{3}^{2}}B.
\end{eqnarray}
There is a scalar gravitational wave with mass $m$ that satisfies the relationship (\ref{relationship between B and C case 2.2}).

\begin{center}
	\textbf{Case 3}: $m^{2} \in \left\{m_{k}^{2}\right\},\det(\mathcal{A})=0$
\end{center}

In this case, Eq. (\ref{scalar mode equation 1 scalar-tensor simply simply simply}) is naturally satisfied. Due to $\det(\mathcal{A})=0$, the set of vectors composed of the first row and the second row of the coefficient matrix $\mathcal{A}$ of Eq. (\ref{scalar mode equation 2 and 3 in matrix form}) is linearly dependent.

For the convenience of the following discussion, we define the following two vectors:
\begin{eqnarray}
	\label{alpha beta}
	\alpha
	&=&
	\left(
	2\Lambda_{2}\prod_{k=1}^{P} \left(m^{2}-{\tilde{m}_{k}}^2\right)^{p_{k}},~
	\Lambda_{4}\prod_{k=1}^{R} \left(m^{2}-{{\hat{m}_{k}}}^2\right)^{r_{k}},~
	2\Lambda_{3}\prod_{k=1}^{Q} \left(m^{2}-{{m^\prime_{k}}}^2\right)^{q_{k}}
	\right),
	\nonumber \\
	\beta
	&=&
	\left(
	-3m^{2}\Lambda_{4}\prod_{k=1}^{R} \left(m^{2}-{{\hat{m}_{k}}}^2\right)^{r_{k}},~
	\Lambda_{5}\prod_{k=1}^{S} \left(m^{2}-{{\overline{m}_{k}}}^2\right)^{s_{k}}, ~
	\Lambda_{4}\prod_{k=1}^{R} \left(m^{2}-{{\hat{m}_{k}}}^2\right)^{r_{k}}
	\right).
\end{eqnarray}
It can be seen that the first two components of the vectors defined above are derived from the coefficient matrix $\mathcal{A}$, and the last component is derived from $\mathcal{B}$.

If the set of $\alpha$ and $\beta$ is linearly independent, then Eq. (\ref{scalar mode equation 2 and 3 in matrix form}) will require $2C-B=0$. In this case, the equation degenerates into the form of Eq. (\ref{scalar mode equation 2 and 3 in matrix form case A}). Therefore, the analysis of scalar gravitational waves in this case is completely similar to the analysis in Case 1.

We can break it down into more specific cases for further discussion.

\textbf{Case 3.1.1}: $m^{2} \in \left\{m_{k}^{2}\right\},\det(\mathcal{A})=0$, the set of $\alpha$ and $\beta$ is linearly independent, $m^{2} \in \left\{\tilde{m}_{k}^{2}\right\}, m^{2} \in \left\{\hat{m}_{k}^{2}\right\}$.  In this case, $B$ and $D$ can take any value. Therefore, there is a scalar gravitational wave with mass $m$ and $C=\frac{B}{2}$.

\textbf{Case 3.1.2}: $m^{2} \in \left\{m_{k}^{2}\right\},\det(\mathcal{A})=0$, the set of $\alpha$ and $\beta$ is linearly independent, $m^{2} \in \left\{\tilde{m}_{k}^{2}\right\}, m^{2} \notin \left\{\hat{m}_{k}^{2}\right\}$. In this case, $D=0$ and there is a scalar gravitational wave with mass $m$ and $C=\frac{B}{2}$.

\textbf{Case 3.1.3}: $m^{2} \in \left\{m_{k}^{2}\right\},\det(\mathcal{A})=0$, the set of $\alpha$ and $\beta$ is linearly independent, $m^{2} \notin \left\{\tilde{m}_{k}^{2}\right\}, m^{2} \in \left\{\hat{m}_{k}^{2}\right\}$. In this case, $B=C=0$ and the theory does not allow scalar gravitational waves.

\textbf{Case 3.1.4}:  $m^{2} \in \left\{m_{k}^{2}\right\},\det(\mathcal{A})=0$, the set of $\alpha$ and $\beta$ is linearly independent, $m^{2} \notin \left\{\tilde{m}_{k}^{2}\right\}, m^{2} \notin \left\{\hat{m}_{k}^{2}\right\}$. In this case,
\begin{eqnarray}
	\label{relationship of B D in case 1.5}
	B=-\frac{\Lambda_{4}\prod_{k=1}^{R} \left(m^{2}-{{\hat{m}_{k}}}^2\right)^{r_{k}}}{2\Lambda_{2}\prod_{k=1}^{P} \left(m^{2}-{\tilde{m}_{k}}^2\right)^{p_{k}}}D.
\end{eqnarray}
Therefore, there is a scalar gravitational wave with mass $m$ and $C=\frac{B}{2}$.

If the set of $\alpha$ and $\beta$ is linearly dependent, then Eqs. (\ref{scalar mode equation 2 scalar-tensor simply simply simply}) and (\ref{scalar mode equation 3 scalar-tensor simply simply simply}) are not independent. At this point, we only need to study Eq. (\ref{scalar mode equation 2 scalar-tensor simply simply simply}).

We can delve into more specific cases for our discussion.

\textbf{Case 3.2.1}: $m^{2} \in \left\{m_{k}^{2}\right\},\det(\mathcal{A})=0$, the set of $\alpha$ and $\beta$ is linearly dependent, $m^{2} \in \left\{\tilde{m}_{k}^{2}\right\}, m^{2} \in \left\{\hat{m}_{k}^{2}\right\}, m^{2} \in \left\{{{m^\prime_{k}}}^2\right\}$. In this case, Eq. (\ref{scalar mode equation 2 scalar-tensor simply simply simply}) is naturally satisfied. Therefore, there is a scalar gravitational wave with mass $m$ and independent $B$ and $C$.

\textbf{Case 3.2.2}: $m^{2} \in \left\{m_{k}^{2}\right\},\det(\mathcal{A})=0$, the set of $\alpha$ and $\beta$ is linearly dependent, $m^{2} \notin \left\{\tilde{m}_{k}^{2}\right\}, m^{2} \in \left\{\hat{m}_{k}^{2}\right\}, m^{2} \in \left\{{{m^\prime_{k}}}^2\right\}$. In this case, Eq. (\ref{scalar mode equation 2 scalar-tensor simply simply simply}) requires $B=0$. Therefore, there is a scalar gravitational wave with mass $m$ and $B=0$.

\textbf{Case 3.2.3}: $m^{2} \in \left\{m_{k}^{2}\right\},\det(\mathcal{A})=0$, the set of $\alpha$ and $\beta$ is linearly dependent, $m^{2} \in \left\{\tilde{m}_{k}^{2}\right\}, m^{2} \notin \left\{\hat{m}_{k}^{2}\right\}, m^{2} \in \left\{{{m^\prime_{k}}}^2\right\}$. In this case, Eq. (\ref{scalar mode equation 2 scalar-tensor simply simply simply}) requires $D=0$. Therefore, there is a scalar gravitational wave with mass $m$ and independent $B$ and $C$.

\textbf{Case 3.2.4}: $m^{2} \in \left\{m_{k}^{2}\right\},\det(\mathcal{A})=0$, the set of $\alpha$ and $\beta$ is linearly dependent, $m^{2} \in \left\{\tilde{m}_{k}^{2}\right\}, m^{2} \in \left\{\hat{m}_{k}^{2}\right\}, m^{2} \notin \left\{{{m^\prime_{k}}}^2\right\}$. In this case, Eq. (\ref{scalar mode equation 2 scalar-tensor simply simply simply}) requires $2C-B=0$. Therefore,  there is a scalar gravitational wave with mass $m$ and $C=\frac{B}{2}$.

\textbf{Case 3.2.5}: $m^{2} \in \left\{m_{k}^{2}\right\},\det(\mathcal{A})=0$, the set of $\alpha$ and $\beta$ is linearly dependent, $m^{2} \notin \left\{\tilde{m}_{k}^{2}\right\}, m^{2} \notin \left\{\hat{m}_{k}^{2}\right\}, m^{2} \in \left\{{{m^\prime_{k}}}^2\right\}$. In this case, Eq. (\ref{scalar mode equation 2 scalar-tensor simply simply simply}) requires
\begin{eqnarray}
	\label{case 3.2.5}
	2\Lambda_{2}\prod_{k=1}^{P} \left(m^{2}-{\tilde{m}_{k}}^2\right)^{p_{k}}B
	+\Lambda_{4}\prod_{k=1}^{R} \left(m^{2}-{{\hat{m}_{k}}}^2\right)^{r_{k}}D
	=
	0.
\end{eqnarray}
Therefore, there is a scalar gravitational wave with mass $m$ and independent $B$ and $C$.

\textbf{Case 3.2.6}: $m^{2} \in \left\{m_{k}^{2}\right\},\det(\mathcal{A})=0$, the set of $\alpha$ and $\beta$ is linearly dependent, $m^{2} \notin \left\{\tilde{m}_{k}^{2}\right\}, m^{2} \in \left\{\hat{m}_{k}^{2}\right\}, m^{2} \notin \left\{{{m^\prime_{k}}}^2\right\}$. In this case, Eq. (\ref{scalar mode equation 2 scalar-tensor simply simply simply}) requires
\begin{eqnarray}
	\label{case 3.2.6}
	2\Lambda_{2}\prod_{k=1}^{P} \left(m^{2}-{\tilde{m}_{k}}^2\right)^{p_{k}}B
	+2k_{3}^{2}\Lambda_{3}\prod_{k=1}^{Q} \left(m^{2}-{{m^\prime_{k}}}^2\right)^{q_{k}}\left(2C-B\right)
	=
	0.
\end{eqnarray}
Therefore, there is a scalar gravitational wave with mass $m$ that satisfies the relationship (\ref{case 3.2.6}).

\textbf{Case 3.2.7}: $m^{2} \in \left\{m_{k}^{2}\right\},\det(\mathcal{A})=0$, the set of $\alpha$ and $\beta$ is linearly dependent, $m^{2} \in \left\{\tilde{m}_{k}^{2}\right\}, m^{2} \notin \left\{\hat{m}_{k}^{2}\right\}, m^{2} \notin \left\{{{m^\prime_{k}}}^2\right\}$. In this case, Eq. (\ref{scalar mode equation 2 scalar-tensor simply simply simply}) requires
\begin{eqnarray}
	\label{case 3.2.7}
	\Lambda_{4}\prod_{k=1}^{R} \left(m^{2}-{{\hat{m}_{k}}}^2\right)^{r_{k}}D
	+2k_{3}^{2}\Lambda_{3}\prod_{k=1}^{Q} \left(m^{2}-{{m^\prime_{k}}}^2\right)^{q_{k}}\left(2C-B\right)
	=
	0.
\end{eqnarray}
Therefore, there is a scalar gravitational wave with mass $m$ and independent $B$ and $C$.

\textbf{Case 3.2.8}: $m^{2} \in \left\{m_{k}^{2}\right\},\det(\mathcal{A})=0$, the set of $\alpha$ and $\beta$ is linearly dependent, $m^{2} \notin \left\{\tilde{m}_{k}^{2}\right\}, m^{2} \notin \left\{\hat{m}_{k}^{2}\right\}, m^{2} \notin \left\{{{m^\prime_{k}}}^2\right\}$. In this case, Eq. (\ref{scalar mode equation 2 scalar-tensor simply simply simply}) requires
\begin{eqnarray}
	2\Lambda_{2}\prod_{k=1}^{P} \left(m^{2}-{\tilde{m}_{k}}^2\right)^{p_{k}}B
	+\Lambda_{4}\prod_{k=1}^{R} \left(m^{2}-{{\hat{m}_{k}}}^2\right)^{r_{k}}D
	\!\!&=&\!\!
	-2k_{3}^{2}\Lambda_{3}\prod_{k=1}^{Q} \left(m^{2}-{{m^\prime_{k}}}^2\right)^{q_{k}}\left(2C-B\right).
	\nonumber\\
	\label{case 3.2.8}
\end{eqnarray}
Therefore, there is a scalar gravitational wave with mass $m$ and independent $B$ and $C$.

As can be seen, as long as $m^{2} \in \left\{m_{k}^{2}\right\}$ or $\det(\mathcal{A})=0$ is satisfied, there exists a non-zero solution given in Eq. (\ref{Thetaphipsi=BCDeikx}). We write the expression for $\det(\mathcal{A})$:
\begin{eqnarray}
	\label{detA}
	\det(\mathcal{A})
	&=&2\Lambda_{2}\Lambda_{5}\prod_{k=1}^{P} \left(m^{2}-{\tilde{m}_{k}}^2\right)^{p_{k}}
	\prod_{k=1}^{S} \left(m^{2}-{{\overline{m}_{k}}}^2\right)^{s_{k}}
	\nonumber \\
	&+&3m^{2}\Lambda_{4}^2\prod_{k=1}^{R} \left(m^{2}-{{\hat{m}_{k}}}^2\right)^{r_{k}}
	\prod_{k=1}^{R} \left(m^{2}-{{\hat{m}_{k}}}^2\right)^{r_{k}}.
\end{eqnarray}
It can be observed that Eq. (\ref{detA}) is a polynomial in $m^{2}$, so $\det(\mathcal{A})$ can always be equivalently written as
\begin{eqnarray}
	\label{detA simply}
	\det(\mathcal{A})
	=
	\Lambda_{6}\prod_{k=1}^{T} \left(m^{2}-{\mathring{m}_{k}}^2\right)^{t_{k}}.
\end{eqnarray}
Therefore, $m^{2} \in \left\{m_{k}^{2}\right\}$ or $\det(\mathcal{A})=0$ is equivalent to $m^{2} \in \left\{m_{k}^{2}\right\} \cup \left\{{\mathring{m}_{k}}^2 \right\}$. To avoid superluminal speed and the linear instability, we also expect each $\mathring{m}_{k}^{2}$ to be a non-negative real number.

Now, we study the polarization modes of the scalar gravitational wave.

For all cases that satisfy $B=0$, using Eq. (\ref{breathing mode and the longitudinal mode}), it can be seen that $P_{6}=0$ and $P_{1}=\left|\left(m^{2}-\omega^{2}\right)C\right|$. The scalar gravitational wave is a longitudinal mode.

For all cases that satisfy $2C-B=0$, using Eq. (\ref{breathing mode and the longitudinal mode}), it can be seen that the scalar gravitational wave is a mixture of the longitudinal mode and the breathing mode, and the amplitude ratio $\mathcal{R}$ of the two modes is ${m^{2}}/{\omega^{2}}$. When $\mathcal{R}=0$, the scalar gravitational wave degenerates into a breathing mode.

For all cases where $B$ and $C$ can take any value, the scalar gravitational wave could be a breathing mode, a longitudinal mode, or a mixed mode of the two.

There are still Case 2.2 and Case 3.2.6.
For Case 2.2, according to Eqs. (\ref{breathing mode and the longitudinal mode}) and (\ref{relationship between B and C case 2.2}), the scalar mode is a mixture of the longitudinal mode and the breathing mode, and the amplitude ratio $\mathcal{R}$ 
is
\begin{eqnarray}
	\label{ratio P1,P6 case2.2}
	\mathcal{R}=
	\left|\frac{P_1}{P_6}\right|=
	\left|
	\frac{1+\gamma m^2}{\gamma \omega^{2}}
	\right|.
\end{eqnarray}
For Case 3.2.6, according to Eqs. (\ref{breathing mode and the longitudinal mode}) and (\ref{case 3.2.6}), the scalar gravitational wave is a mixture of the longitudinal mode of the breathing mode, and their amplitude ratio $\mathcal{R}$
is
\begin{eqnarray}
	\label{ratio P1,P6 case3.2.6}
	\mathcal{R}=
	\left|\frac{P_1}{P_6}\right|=
	\left|
	\frac{m^{2}\Lambda_{3}\prod_{k=1}^{Q} \left(m^{2}-{{m^\prime_{k}}}^2\right)^{q_{k}}
	-\Lambda_{2}\prod_{k=1}^{P} \left(m^{2}-{\tilde{m}_{k}}^2\right)^{p_{k}}}
	{\omega^{2}\Lambda_{3}\prod_{k=1}^{Q} \left(m^{2}-{{m^\prime_{k}}}^2\right)^{q_{k}}}
	\right|.
\end{eqnarray}

We summarize the results of scalar gravitational waves in Tab. \ref{tab:scalar mode in general scalar-tensor theory}.

\begin{center}
	\begin{table}[htbp]
		\resizebox{\textwidth}{!}{
			\begin{tabular}{|c|l|c|c|c|c|}
				\hline
				\hline
				\textbf{Cases} & \qquad\textbf{Conditions} & Breathing mode & Longitudinal mode & Dependent or not & $\mathcal{R}=m^2/{\omega}^2$ \\
				\hline
				case 1.1 &$m^{2} \notin \left\{m_{k}^{2}\right\},\det(\mathcal{A}) \neq 0$& $\times$ & $\times$ & - & - \\
				\hline
				case 1.2 &$m^{2} \notin \left\{m_{k}^{2}\right\},\det(\mathcal{A})=0, m^{2} \in \left\{\tilde{m}_{k}^{2}\right\}, m^{2} \in \left\{\hat{m}_{k}^{2}\right\}$& $\checkmark$ & $\checkmark$ & $\checkmark$ & $\checkmark$ \\
				\hline
				case 1.3 &$m^{2} \notin \left\{m_{k}^{2}\right\},\det(\mathcal{A})=0, m^{2} \in \left\{\tilde{m}_{k}^{2}\right\}, m^{2} \notin \left\{\hat{m}_{k}^{2}\right\}$& $\checkmark$ & $\checkmark$ & $\checkmark$ & $\checkmark$ \\
				\hline
				case 1.4 &$m^{2} \notin \left\{m_{k}^{2}\right\},\det(\mathcal{A})=0, m^{2} \notin \left\{\tilde{m}_{k}^{2}\right\}, m^{2} \in \left\{\hat{m}_{k}^{2}\right\}$& $\times$ & $\times$ & - & - \\
				\hline
				case 1.5 &$m^{2} \notin \left\{m_{k}^{2}\right\},\det(\mathcal{A})=0, m^{2} \notin \left\{\tilde{m}_{k}^{2}\right\}, m^{2} \notin \left\{\hat{m}_{k}^{2}\right\}$& $\checkmark$ & $\checkmark$ & $\checkmark$ & $\checkmark$ \\
				\hline
				case 2.1 &$m^{2} \in \left\{m_{k}^{2}\right\},\det(\mathcal{A}) \neq 0, \gamma=0$& $\times$ & $\checkmark$ & - & - \\
				\hline
				case 2.2 &$m^{2} \in \left\{m_{k}^{2}\right\},\det(\mathcal{A}) \neq 0, \gamma \neq 0$& $\checkmark$ & $\checkmark$ & $\checkmark$ & $\times$ \\
				\hline
				case 3.1.1 &$m^{2} \in \left\{m_{k}^{2}\right\},\det(\mathcal{A})=0$, $\alpha$, $\beta$ linearly independent, $m^{2} \in \left\{\tilde{m}_{k}^{2}\right\}, m^{2} \in \left\{\hat{m}_{k}^{2}\right\}$& $\checkmark$ & $\checkmark$ & $\checkmark$ & $\checkmark$ \\
				\hline
				case 3.1.2 &$m^{2} \in \left\{m_{k}^{2}\right\},\det(\mathcal{A})=0$, $\alpha$, $\beta$ linearly independent, $m^{2} \in \left\{\tilde{m}_{k}^{2}\right\}, m^{2} \notin \left\{\hat{m}_{k}^{2}\right\}$& $\checkmark$ & $\checkmark$ & $\checkmark$ & $\checkmark$ \\
				\hline
				case 3.1.3 &$m^{2} \in \left\{m_{k}^{2}\right\},\det(\mathcal{A})=0$, $\alpha$, $\beta$ linearly independent, $m^{2} \notin \left\{\tilde{m}_{k}^{2}\right\}, m^{2} \in \left\{\hat{m}_{k}^{2}\right\}$& $\times$ & $\times$ & - & - \\
				\hline
				case 3.1.4 & $m^{2} \in \left\{m_{k}^{2}\right\},\det(\mathcal{A})=0$, $\alpha$, $\beta$ linearly independent, $m^{2} \notin \left\{\tilde{m}_{k}^{2}\right\}, m^{2} \notin \left\{\hat{m}_{k}^{2}\right\}$& $\checkmark$ & $\checkmark$ & $\checkmark$ & $\checkmark$ \\
				\hline
				case 3.2.1 &$m^{2} \in \left\{m_{k}^{2}\right\},\det(\mathcal{A})=0$, $\alpha$, $\beta$ linearly dependent, $m^{2} \in \left\{\tilde{m}_{k}^{2}\right\}, m^{2} \in \left\{\hat{m}_{k}^{2}\right\}, m^{2} \in \left\{{{m^\prime_{k}}}^2\right\}$& $\checkmark$ & $\checkmark$ & $\times$ & - \\
				\hline
				case 3.2.2 &$m^{2} \in \left\{m_{k}^{2}\right\},\det(\mathcal{A})=0$, $\alpha$, $\beta$ linearly dependent, $m^{2} \notin \left\{\tilde{m}_{k}^{2}\right\}, m^{2} \in \left\{\hat{m}_{k}^{2}\right\}, m^{2} \in \left\{{{m^\prime_{k}}}^2\right\}$& $\times$ & $\checkmark$ & - & - \\
				\hline
				case 3.2.3 &$m^{2} \in \left\{m_{k}^{2}\right\},\det(\mathcal{A})=0$, $\alpha$, $\beta$ linearly dependent, $m^{2} \in \left\{\tilde{m}_{k}^{2}\right\}, m^{2} \notin \left\{\hat{m}_{k}^{2}\right\}, m^{2} \in \left\{{{m^\prime_{k}}}^2\right\}$& $\checkmark$ & $\checkmark$ & $\times$ & - \\
				\hline
				case 3.2.4 &$m^{2} \in \left\{m_{k}^{2}\right\},\det(\mathcal{A})=0$, $\alpha$, $\beta$ linearly dependent, $m^{2} \in \left\{\tilde{m}_{k}^{2}\right\}, m^{2} \in \left\{\hat{m}_{k}^{2}\right\}, m^{2} \notin \left\{{{m^\prime_{k}}}^2\right\}$& $\checkmark$ & $\checkmark$ & $\checkmark$ & $\checkmark$ \\
				\hline
				case 3.2.5 &$m^{2} \in \left\{m_{k}^{2}\right\},\det(\mathcal{A})=0$, $\alpha$, $\beta$ linearly dependent, $m^{2} \notin \left\{\tilde{m}_{k}^{2}\right\}, m^{2} \notin \left\{\hat{m}_{k}^{2}\right\}, m^{2} \in \left\{{{m^\prime_{k}}}^2\right\}$& $\checkmark$ & $\checkmark$ & $\times$ & - \\
				\hline
				case 3.2.6 &$m^{2} \in \left\{m_{k}^{2}\right\},\det(\mathcal{A})=0$, $\alpha$, $\beta$ linearly dependent, $m^{2} \notin \left\{\tilde{m}_{k}^{2}\right\}, m^{2} \in \left\{\hat{m}_{k}^{2}\right\}, m^{2} \notin \left\{{{m^\prime_{k}}}^2\right\}$& $\checkmark$ & $\checkmark$ & $\checkmark$ & $\times$ \\
				\hline
				case 3.2.7 &$m^{2} \in \left\{m_{k}^{2}\right\},\det(\mathcal{A})=0$, $\alpha$, $\beta$ linearly dependent, $m^{2} \in \left\{\tilde{m}_{k}^{2}\right\}, m^{2} \notin \left\{\hat{m}_{k}^{2}\right\}, m^{2} \notin \left\{{{m^\prime_{k}}}^2\right\}$& $\checkmark$ & $\checkmark$ & $\times$ & - \\
				\hline
				case 3.2.8 &$m^{2} \in \left\{m_{k}^{2}\right\},\det(\mathcal{A})=0$, $\alpha$, $\beta$ linearly dependent, $m^{2} \notin \left\{\tilde{m}_{k}^{2}\right\}, m^{2} \notin \left\{\hat{m}_{k}^{2}\right\}, m^{2} \notin \left\{{{m^\prime_{k}}}^2\right\}$& $\checkmark$ & $\checkmark$ & $\times$ & - \\
				\hline
				\hline
		\end{tabular}  }
		\caption{Scalar modes in various cases in the most general scalar-tensor theory.}
		\label{tab:scalar mode in general scalar-tensor theory}
	\end{table}
\end{center}

Now, we have completed the study of polarizations in the most general scalar-tensor theory.

Finally, it should be emphasized that the most general scalar-tensor theory still satisfies the four universal propositions given in Sec. \ref{sec: 5}. This is because the proof of these four propositions only requires Eqs. (\ref{tensor mode equation of general metric theory}), (\ref{vector mode equation metric}), and (\ref{scalar mode equation 1 metric simply}). These equations also hold true in the most general scalar-tensor theory.

\section{Conclusion}
\label{sec: 7}
In this paper, we have formulated the most general second-order actions for perturbations in both the most general metric theory and the most general scalar-tensor theory, satisfying the four conditions outlined in the introduction. We also have some implicit assumptions: (1) The existence of a flat spacetime background; (2) The theory is an extension of general relativity; (3) The Lagrangian is analytical. Then we studied the polarization modes of gravitational waves in the general metric theory and the general scalar-tensor theory. We found that both the most general metric theory and scalar-tensor theory allow for up to all six polarization modes, and these modes depend on the parameter space. The specific properties of the polarization modes and  the speed of gravitational waves in the two theories under different parameter spaces were listed in Sec. \ref{sec: 4} and Sec. \ref{sec: 7}, respectively. Therefore, the study of the polarization modes and speed of gravitational waves in the specific general metric theory and  scalar-tensor theory will be entirely attributed to the determination of the parameters in the second-order action given in this paper.

We also found that whether it is the general metric theory or the general scalar-tensor theory, the polarization modes satisfy the following four properties: (1) If the theory allows vector gravitational waves with mass $m$, then there must be tensor gravitational waves with the same mass $m$ in the theory. Correspondingly, if the theory allows tensor gravitational waves with mass $m$ nonvanishing, then there must be vector gravitational waves with the same mass $m$. (2) If the general metric theory allows tensor gravitational waves propagating only at the speed of light, then there must be no vector gravitational waves. (3) If the theory allows tensor gravitational waves propagating only at the speed of light, then for a scalar gravitational wave with mass $m$ and frequency $\omega$, it must be a mixed mode of the longitudinal mode and breathing mode. And the amplitude ratio of the two modes must be $m^2/{\omega}^2$ (Therefore, when $m=0$, it degenerates into a breathing mode). (4) If the amplitude ratio of the longitudinal mode and the breathing one of a scalar gravitational wave with mass $m$ and frequency $\omega$ is not $m^2/{\omega}^2$, then the theory must have a field equation higher than the second derivative and must have tensor and vector gravitational waves with speed less than the speed of light. The existence of these universal properties indicates that the detection of polarization modes of gravitational waves can be used not only to test various modified gravitational theories but also to test basic physical principles. If not all of these properties are found to be correct, then either the gravity theory that describes our world is not the general metric theory or the general scalar-tensor theory, or we need to modify at least one fundamental physical principle.

We have studied polarization modes of gravitational waves in the general metric theory and scalar-tensor theory. An interesting question is whether this research method can be extended to other modified gravity theories, such as the vector-tensor theory and the scalar-vector-tensor theory. It is necessary to conduct similar research on the polarization modes of gravitational waves in other modified gravitational theories and verify whether the universal properties we have identified remain valid. This aspect will be left for future work.

\section*{Acknowledgments}
This work is supported in part by the National Key Research and Development Program of China (Grant No. 2020YFC2201503), the National Natural Science Foundation of China (Grants No. 11875151 and No. 12247101), the 111 Project (Grant No. B20063), the Department of
education of Gansu Province: Outstanding Graduate ``Innovation Star" Project (Grant
No. 2023CXZX-057), the Major Science and Technology Projects of Gansu Province,  and ``Lanzhou City's scientic research funding subsidy to Lanzhou University".


\end{document}